\documentclass[12pt]{article}
\usepackage{jheppub}
\pdfoutput=1
\usepackage{epstopdf}

\usepackage{jheppub}
\usepackage{mathrsfs}
\usepackage{psfrag}
\usepackage{color}

\usepackage[table]{xcolor}

\usepackage{mathrsfs,amsmath,bbm,array,amsfonts,graphicx,wrapfig,arydshln,lscape,float,multirow,longtable,rotating,subfigure,makecell}
\usepackage{url}
\usepackage[utf8]{inputenc}

\newcommand{\be}{\begin{equation}}
\newcommand{\ee}{\end{equation}}
\newcommand{\beq}{\begin{equation}}
\newcommand{\beql}[1]{\begin{equation}\label{#1}}
\newcommand{\eeq}{\end{equation}}
\newcommand{\ba}{\begin{array}}
\newcommand{\ea}{\end{array}}
\newcommand{\bea}{\begin{eqnarray}}
\newcommand{\beal}[1]{\begin{eqnarray}\label{#1}}
\newcommand{\eea}{\end{eqnarray}}
\newcommand{\ben}{\begin{enumerate}}
\newcommand{\een}{\end{enumerate}}
\newcommand{\bean}{\begin{eqnarray*}}
\newcommand{\eean}{\end{eqnarray*}}
\newcommand{\eref}[1]{(\ref{#1})}
\newcommand{\sref}[1]{\S\ref{#1}}

\newcommand{\fref}[1]{Figure \ref{#1}}
\newcommand{\btab}[1]{\begin{tabular}{#1}}
\newcommand{\etab}{\end{tabular}}

\newcommand{\comment}[1]{}

\newcommand{\qed}{\nobreak \ifvmode \relax \else
      \ifdim\lastskip<1.5em \hskip-\lastskip
      \hskip1.5em plus0em minus0.5em \fi \nobreak
      \vrule height0.75em width0.5em depth0.25em\fi}

\newcolumntype{C}[1]{>{\centering\arraybackslash}m{#1}}

\newcommand{\pl}{Pl\"ucker }

\newcommand{\incell}[1]{\mathtt{\displaystyle \color[RGB]{69,78,153} In[#1]:=} \; \;}
\newcommand{\outcell}[1]{\mathtt{\displaystyle \color[RGB]{69,78,153} Out[#1]:=} \; \;}
\newcommand{\func}[2]{\text{\tt {\color[RGB]{0,44,195} #1}[}#2\text{\tt ]}}
\newcommand{\varbar}[1]{\text{\tt{\color[RGB]{67,137,88}{\sl #1\_}}}}
\newcommand{\varbardef}[2]{\text{\tt{\color[RGB]{67,137,88}{\sl #1\_:}}{\tt #2}}}
\newcommand{\var}[1]{\text{\tt{\color[RGB]{67,137,88}{\sl #1}}}}
\newcommand{\pc}[1]{\text{\tt #1}}
\newcommand{\N}[1]{$\mathcal{N}=#1$}
\newcommand{\kast}{\varbar{topleft},\varbar{topright},\varbar{bottomleft},\varbar{bottomright}}
\newcommand{\point}{\vspace{0.3cm} \noindent {\color[RGB]{0,0,0} $\bullet$} \hspace{1pt} }
\newcommand{\pointscat}{\vspace{0.3cm} \noindent {\color[RGB]{255,51,51} $\bullet$} \hspace{1pt} }
\newcommand{\pointbft}{\vspace{0.3cm} \noindent {\color[RGB]{180,0,180} $\bullet$} \hspace{1pt} }

\title{Bipartite Graphs, On-Shell Diagrams, and Bipartite Field Theories: a Computational Package in Mathematica}

\author{Daniele Galloni}

\affiliation{INFN, Sezione di Torino \\
Via Pietro Giuria 1, 10125 Torino, Italy
}

\emailAdd{daniele.galloni@to.infn.it}

\abstract{We present \pc{bipartiteSUSY}, a Mathematica package designed to perform calculations for physical theories based on bipartite graphs. In particular, the package can employ the recently developed arsenal of techniques surrounding on-shell diagrams in \N{4} SYM scattering amplitudes, including those for non-planar diagrams, with particular attention to computational speed. It also contains a host of tools for computations in \N{1} Bipartite Field Theories, which utilize the same bipartite graphs. Through the use of an interactive graphical tool, it is possible to draw the desired diagrams on the screen and compute commonly sought-after features. The package should be easily accessible to users with little or no previous experience in dealing with bipartite graphs and their combinatorial descriptions.
}

\preprint{
}

\begin{document}

\maketitle

\section{Introduction} 
\label{sec:intro}

Reformulating theories in terms of simple objects that can be intuitively understood plays an important role in their development. While not formally necessary, the rephrasing of quantities into new, often graphical, forms leaves us with an uncluttered and more transparent view of the physics behind the theory in question; moreover, it allows us to detect graphical patterns which were completely obscure in their corresponding mathematical expressions.

Along these lines, the past two decades have enjoyed exceptional progress, where prominent examples of graphical tools are particularly abundant in supersymmetric theories. In \cite{Benvenuti:2004dy,Franco:2005rj,Franco:2005sm,Butti:2005sw,Hanany:2005ss,Feng:2005gw,Franco:2012mm,Franco:2012wv} it was found that a large set of infinite classes of four-dimensional \N{1} theories, first known as \textit{dimers} but later extended to the more general \textit{Bipartite Field Theories} (BFTs), have a very concise graphical representation in terms of bipartite graphs. Moreover, complicated field-theoretic operations have counterparts in the form of simple graphical operations. From a graph-combinatorial perspective, these operations were natural actions to perform on the graph in the first place, a realization which highlighted a deep connection between physical quantities in such theories with the combinatorics of bipartite graphs.

Another important realization of this principle arises in scattering amplitudes in \N{4} SYM. In particular in the planar sector there is enough structure to support very powerful mathematical constructs; it is likely that we have only scratched the surface in our understanding of this theory. In \cite{ArkaniHamed:2012nw} it was shown how planar scattering amplitudes, to all loop orders, can be completely expressed as a sum of bipartite diagrams known as \textit{on-shell diagrams}.\footnote{Strictly speaking, the construction of on-shell diagrams makes them bicolored and trivalent; there is however a trivial graphical operation that transforms these into bipartite graphs containing nodes of arbitrary valency.} These diagrams are individually invariant under the full symmetry group of the theory, transparently display an intimate connection with the Grassmannian \cite{2006math09764P}, and trivially exhibit the surprising \textit{dlog} structure of the amplitude integrand.\footnote{Interesting related progress for $\mathcal{N}<4$ can be found in \cite{Benincasa:2015zna,Benincasa:2016awv}.}

As already stressed above, while a graphical avatar of physical quantities is not formally required, it is precisely this visual formulation that provides a shortcut to our understanding of the theory. Indeed, recent advances in non-planar on-shell diagrams \cite{Franco:2015rma, Chen:2015qna, Badger:2015lda, Bern:2015ple, Frassek:2016wlg, Bourjaily:2016mnp} rest precisely on this principle---if planar on-shell diagrams concisely describe planar scattering amplitudes, it is natural to wonder whether non-planar on-shell diagrams can describe non-planar amplitudes in a similar way. There is increasing evidence suggesting that the answer is affirmative;\footnote{See e.g.\ \cite{Bern:1997nh,Bern:2007hh,Bern:2010tq,Carrasco:2011mn,Bern:2012uc,Arkani-Hamed:2014via,Arkani-Hamed:2014bca,Bern:2014kca,Franco:2015rma,Chen:2015bnt,Badger:2015lda,Bern:2015ple, Frassek:2016wlg, Heslop:2016plj, Herrmann:2016qea} for a discussion on these and closely related topics.} if so, this may be a sign that there is additional hidden structure in the theory yet to be discovered \cite{Bern:2015ple}.

It is interesting and surprising that on-shell diagrams are based on the same graphs that describe \N{1} BFTs. It is even more surprising that on-shell diagrams are subject to the same graphical operations and invariances of BFTs, and that the (different) physical quantities which they represent are encoded by the same graph-combinatorial objects. Without speculating on possible physical connections between the two uses of bipartite graphs, it is clear that discoveries made in one scenario are likely to spill over into realizations for the other scenario, and vice-versa. Also, the graphical tools developed in one scenario are very likely to have some meaning in the other. This was indeed found to be the case and put to use in \cite{Franco:2013nwa,Franco:2014nca,Franco:2015rma,Bourjaily:2016mnp}.

The huge progress described above has left a trail of techniques, many of which are highly graphical in nature, developed for the efficient computation of quantities in these theories. The aim of this paper and the Mathematica package that comes with it is to systematize these tools into a single framework, aimed at intuitive simplicity (and, not least, computational speed). This should make recent advancements in these fields more widely accessible, as well as expedite future exploration of theories representable by bipartite graphs. Among other things, the developed Mathematica functions calculate:
\smallskip 
\begin{itemize}
\item The Grassmannian matrix, its dimension and the reducibility of any planar or non-planar bipartite graph.
\item The full singularity structure of any bipartite graph and its Euler number.
\item Graphical manipulations such as collapsing bivalent nodes, square moves, and bubble reduction, along with interactive and static graph-drawing applications.
\item The planarity of a graph, where ``planar'' in this context means that there is an embedding such that all external edges lie on a disk and no edges cross.
\item Loop-variable degrees of freedom of the graph.
\item Interpreting the graphs as \N{1} BFTs, it also computes zig-zag paths, moduli spaces, and full graphical reductions.
\end{itemize}
\smallskip
In the package, whenever it is relevant, it is possible to explicitly specify whether a given function is applied in a \N{4} scattering amplitudes context or whether it is in a \N{1} BFT context.\footnote{There are two types of BFTs for a given graph, which differ by which symmetry groups of the theory are gauge groups \cite{Franco:2012wv}. These are known as BFT$_1$s, which depend on the surface on which the graph is embedded, and BFT$_2$s, which are independent of the surface embedding and only depend on the connectivity of the graph. Since \N{4} on-shell diagrams are also independent of embedding, the functions in this package often have as default the embedding-independent computation of quantities.}

\bigskip

\section{Getting started} 
\label{sec:gettingstarted}

Before illustrating the functionality of the package we'll briefly explain how to get it working in a Mathematica notebook. Setting up the package is very simple. From the arXiv preprint server the relevant files may be downloaded with the source files for this paper.\footnote{This is done as follows: begin at the abstract page for this paper; click on ``Other formats''; click to download the Source files. It may be that the files are downloaded without file extension, in which case it will be necessary to tag onto the end of the downloaded file's name \pc{.tar.gz}. The file will then need uncompressing.} 

The package is the file \pc{bipartiteSUSY.m}. To use the package, it is sufficient to place this file in a convenient directory; then, in any notebook that has been saved in that same directory, the package is loaded by evaluating
\begin{flalign*}
& \quad \quad \quad \incell{1} \pc{SetDirectory[NotebookDirectory[]];} & \\
& \quad \quad \quad \quad \quad \quad \quad \pc{<<bipartiteSUSY.m} &
\end{flalign*}
which will give the user access to all of the functionality of the package.

It is also possible avoid the need for placing the package in the same directory, by installing the package in such a way that it is made available to all Mathematica notebooks. To do this, simply open any Mathematica notebook, go to ``File'', then ``Install'', then set the type of item to install to ``Package'' and finally select the Source to ``From file'' and choose the file \pc{bipartiteSUSY.m}. After installing, the package can be loaded in any future Mathematica session by evaluating
\begin{flalign*}
& \quad \quad \quad \incell{1} \pc{Needs["bipartiteSUSY`"];} & 
\end{flalign*}
(note that the final character is a tick, not an apostrophe), or 
\begin{flalign*}
& \quad \quad \quad \incell{1} \pc{<<bipartiteSUSY`;} & 
\end{flalign*}
If the package is installed as described above, the downloaded package file is stored in Mathematica's directories and may hence be removed from the downloaded location.

\subsection{\pc{bipartiteSUSY}'s Interactive Graph Drawing} 
\label{sec:drawgraph}

Nearly all functions in this package have a single, simple starting point: an adjacency matrix known as the \textit{Kasteleyn matrix}, which contains the full information of the diagram in question. We shall always assume all diagrams to be bipartite rather than simply bicolored; there are trivial operations for achieving this \cite{ArkaniHamed:2012nw}. 

Before we introduce the Kasteleyn matrix, we shall begin by illustrating a convenient interactive graph-drawing tool provided in the package, which can automatically generate the required Kasteleyn and apply some commonly-used functions to the diagram. After loading the package in a notebook, evaluate
\begin{flalign*}
& \quad \quad \quad \incell{2} \pc{drawGraph[]} & 
\end{flalign*}
without semicolon at the end of the command. This will output the following interactive graph-drawing box\footnote{The precise look of the box will depend on the version of Mathematica and the operating system used.}
\begin{figure}[hbt]
\begin{center}
\includegraphics[scale=0.6]{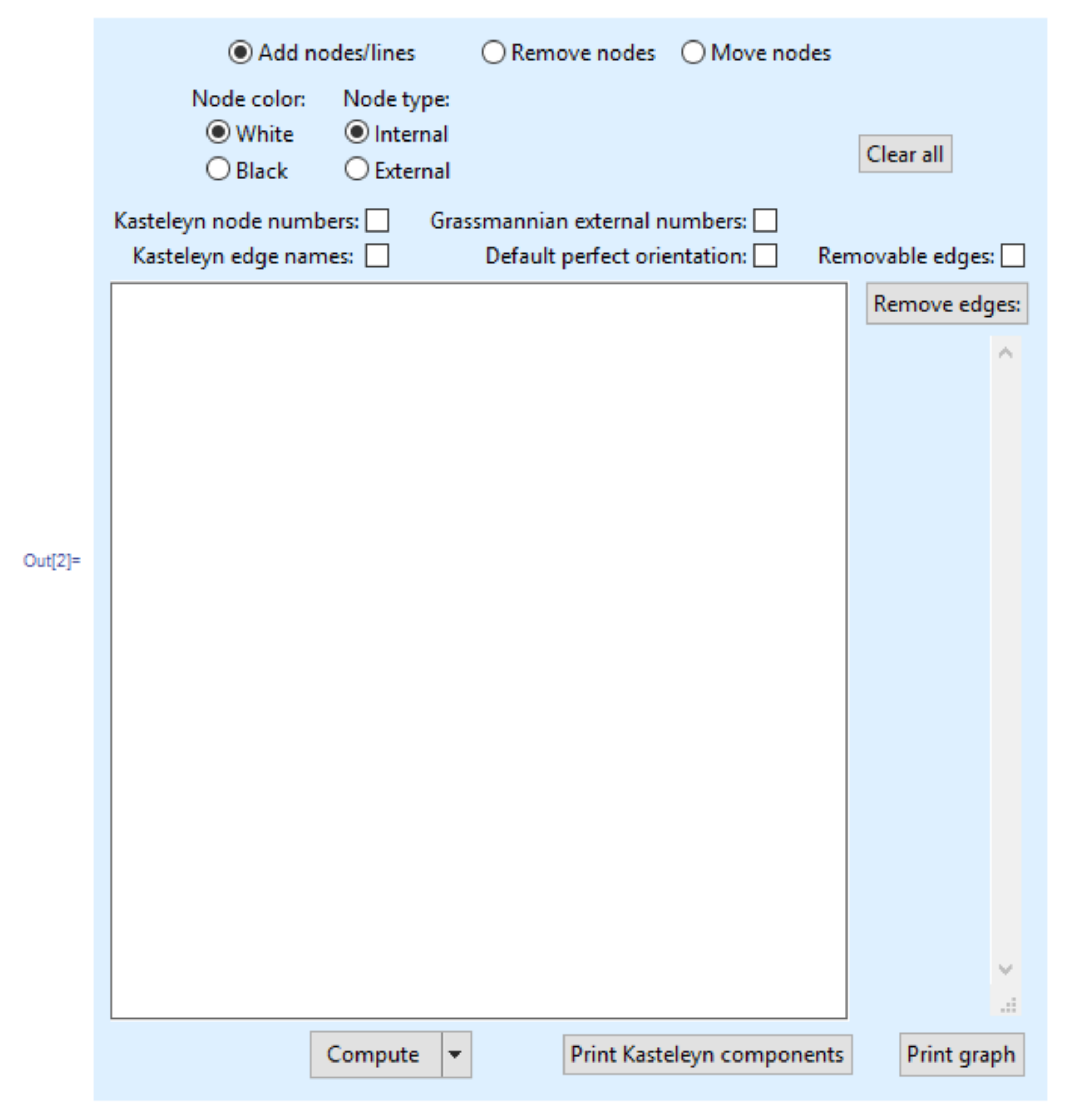}
\label{fig:drawGraphEmpty}
\end{center}
\end{figure}

\paragraph{Drawing a graph.}
Drawing a graph is done by simply clicking and dragging with the mouse inside the white area. The program will create the first node according to the color specified in the options above the white area, and subsequently attempt to alternate the colors so as to create a bipartite object. Using the options, it is also possible to specify whether the next created node is ``external'', meaning the endpoint of an external edge, or ``internal''. External nodes are drawn with a red border to differentiate them from the internal ones. For convenience, clicking on a previously drawn node will swap its color. Right-clicking on a node will change it from internal to external, and vice-versa.

Clicking without dragging on a white area will simply create an isolated node of the specified color and type. In this way it is also possible to create a graph by first placing all the nodes, and subsequently dragging lines between the nodes to connect them.

\paragraph{Modifying a graph.}
Once the graph has been drawn, there are two modifications we can make: moving nodes, or removing nodes and edges. 
\begin{itemize}
\item To remove a node, specify ``Remove nodes'' from the options bar and click on the node that is to be removed. This will also remove all edges connecting to it. To remove individual edges, click on the ``Remove edges button''; this will produce a list of edges in the graph, labeled by which node-numbers they connect. To see the node numbers printed on the graph, simply tick the ``Kasteleyn node numbers'' box. To remove an edge, click on the edge name in the list of edges. Finally, to remove \textit{all} nodes and edges and effectively start over with the graph, click on the ``Clear all'' button.
\item Sometimes a graph may be simple but be drawn in a complicated way, with an excessive number of edges crossing. To move nodes to more convenient locations, select ``Move nodes'' from the options bar. To move a node, simply click and drag it to a new location.
\end{itemize}
Once the graph drawing is complete, there is the possibility of printing out the graph in the notebook, by clicking on the ``Print graph'' button.

\paragraph{Displaying various properties of a graph.}
It is additionally possible to decorate the graph with labels and highlights by using the tick-boxes above the white area. \textbf{\textit{Note:}} The following labels refer to the Kasteleyn-matrix components that are \textit{outputted} by \pc{drawGraph}. As we describe at the end of \sref{sec:kasteleynintro}, it is also possible to give \pc{drawGraph} a Kasteleyn matrix as \textit{input}, and have the function automatically draw the graph corresponding to the matrix. In this case, the labels drawn by the tick-boxes described below will in general be completely unrelated to the notation of this input Kasteleyn matrix.
\begin{itemize}
\setlength\itemsep{0pt}
\item[$\square$] \pc{Kasteleyn node numbers}. As we shall describe below, the graph is associated with the Kasteleyn adjacency matrix, which is the starting point for most of the functions in this package. To know how the nodes in the Kasteleyn matrix correspond to the nodes of the graph, it is possible to tick the ``Kasteleyn node numbers'' box, which will display a small blue node number next to each node. 
\item[$\square$] \pc{Kasteleyn edge names}. Each edge in the Kasteleyn matrix bears a different name; the box ``Kasteleyn edge names'' displays in the graph the edge names used in the Kasteleyn matrix.
\item[$\square$] \pc{Grassmannian external numbers}. \pc{drawGraph} is able to immediately generate the path matrix associated to a drawn on-shell diagram. To know which external nodes correspond to the various columns of the path matrix, it is possible to tick the box ``Grassmannian external numbers'', which will draw green numbers next to the external nodes.
\item[$\square$] \pc{Default perfect orientation}. To view the perfect orientation used in the generated path matrix, the box ``Default perfect orientation'' will draw arrows on the edges of the graph.
\item[$\square$] \pc{Removable edges}. It is often useful to view directly on the graph which edges are \textit{removable}, i.e.\ which edges will yield a reduced graph once they've been removed. Ticking ``Removable edges'' will highlight those edges in orange. Since the computation of the removable edges may take a long time for large or complicated graphs, this computation is done in the kernel, as is further discussed in \sref{sec:finalwords}. A positive consequence of this is that it is possible to stop the computation by aborting it in the usual way (e.g.\ by pressing \pc{Alt}+\pc{.}). 

Keeping the box ticked while drawing will update the removable edges as the graph changes; however, to avoid delaying the creation of the graph while computing its removable edges, the box will automatically untick if the computation of the removable edges exceeds 1 second. Since this may go unnoticed, a warning note appears the first time ``Removable edges'' is ticked, to remind users of this feature.
\end{itemize}

\paragraph{Evaluating functions on a graph.}
In its current form, the graph is embedding independent; in \sref{sec:kasteleynintro} we will discuss how to specify the embedding of the graph in the case of BFTs. There are a number of commonly used functions which can be directly computed on the graph, in this case interpreted as a \N{4} on-shell diagram: the program can compute the reducibility of the diagram; output its path matrix (with an automatically chosen perfect orientation); give the dimension of the corresponding element of the Grassmannian; give the number of boundaries and Euler number of the stratification of the diagram (i.e.\ its singularity structure); output whether the on-shell form associated to the diagram has poles that are not simple \pl coordinates; give the minors of the path matrix written in a gauge-independent way as a sum of perfect matchings; output the planarity of the diagram; compute the helicity $k$ of the associated N$^{k-2}$MHV process; and list all perfect matchings.

All these features are accessed by clicking on the ``Compute'' button under the graph and selecting the required computation; the output will then appear printed under the blue graph-drawing toolbox. Some features are computationally heavy and may take a minute or more to compute, such as the numbers of boundaries in the stratification.

\paragraph{Creating the Kasteleyn for a graph.}
Finally, once the graph is complete, the feature that is perhaps most useful is the ``Print Kasteleyn components'' button. This prints out the Kasteleyn matrix, which unlocks essentially all functionality of the Mathematica package. We shall now move on to briefly discussing what this matrix is.

\paragraph{\textit{Note}.}
Opening a saved notebook containing dynamic content from \pc{drawGraph} will prompt the user with an automatic warning. This warning can be safely ignored in this case; it is presumably generated due to generic security concerns with dynamic content.

\subsection{The Kasteleyn Matrix} 
\label{sec:kasteleynintro}

The Kasteleyn matrix is a weighted adjacency matrix for bipartite graphs. Moreover, by construction, it differentiates between internal and external edges and nodes---this structure allows for the use of very powerful combinatorial tools to compute the quantities of interest.\footnote{The external nodes have no physical meaning. They are however useful in order to treat the diagram using a unified framework; therefore, in this package, it will always be necessary to decorate the diagram with external nodes at the endpoints of external edges.} We shall here briefly describe the structure of the Kasteleyn matrix; for a more detailed discussion (in particular in the context of BFT$_1$s) the reader is encouraged to read \cite{Franco:2012mm}.

The Kasteleyn is constructed by first assigning an individual weight to each edge in the graph. Since the graph is bipartite, white nodes will only connect to black nodes and vice-versa. We may thus construct a matrix with rows representing white nodes, columns representing black nodes, and entries in the matrix given by the sum of edges\footnote{Here and in what follows we will often write \textit{edges} to mean \textit{edge weights}. We hope the reader will not be confused by this slight abuse of terminology.} between any given white and black node. If we further separate those rows associated to internal nodes from those associated to external nodes, and do the same for the columns, we obtain the Kasteleyn matrix, which has the schematic structure
\begin{equation} \label{eq:kasteleynstructure}
K = \left(\begin{array}{c|c|c} & \ \ B_i \ \ & \ \ B_e \ \ \\ \hline
W_i & * & * \\ \hline
W_e & * & 0 \\
\end{array}
\right) \; ,
\end{equation}
where $W_i$ and $B_i$ represent internal white and black nodes, respectively, and $W_e$ and $B_e$ represent the external white and black nodes. Since external nodes by definition only connect to internal nodes, the bottom-right part of the Kasteleyn matrix will only contain zeros. The number of white and black nodes may be different, so the Kasteleyn will not necessarily be square.

\paragraph{\textit{Example}.}
We shall go through a simple example that illustrates how straightforward it is to construct the Kasteleyn matrix, which in turn gives access to all of the package's functionality. Using the \pc{drawGraph[]} function, we draw a relatively simple diagram:
\begin{figure}[hbt]
\begin{center}
\includegraphics[scale=0.6]{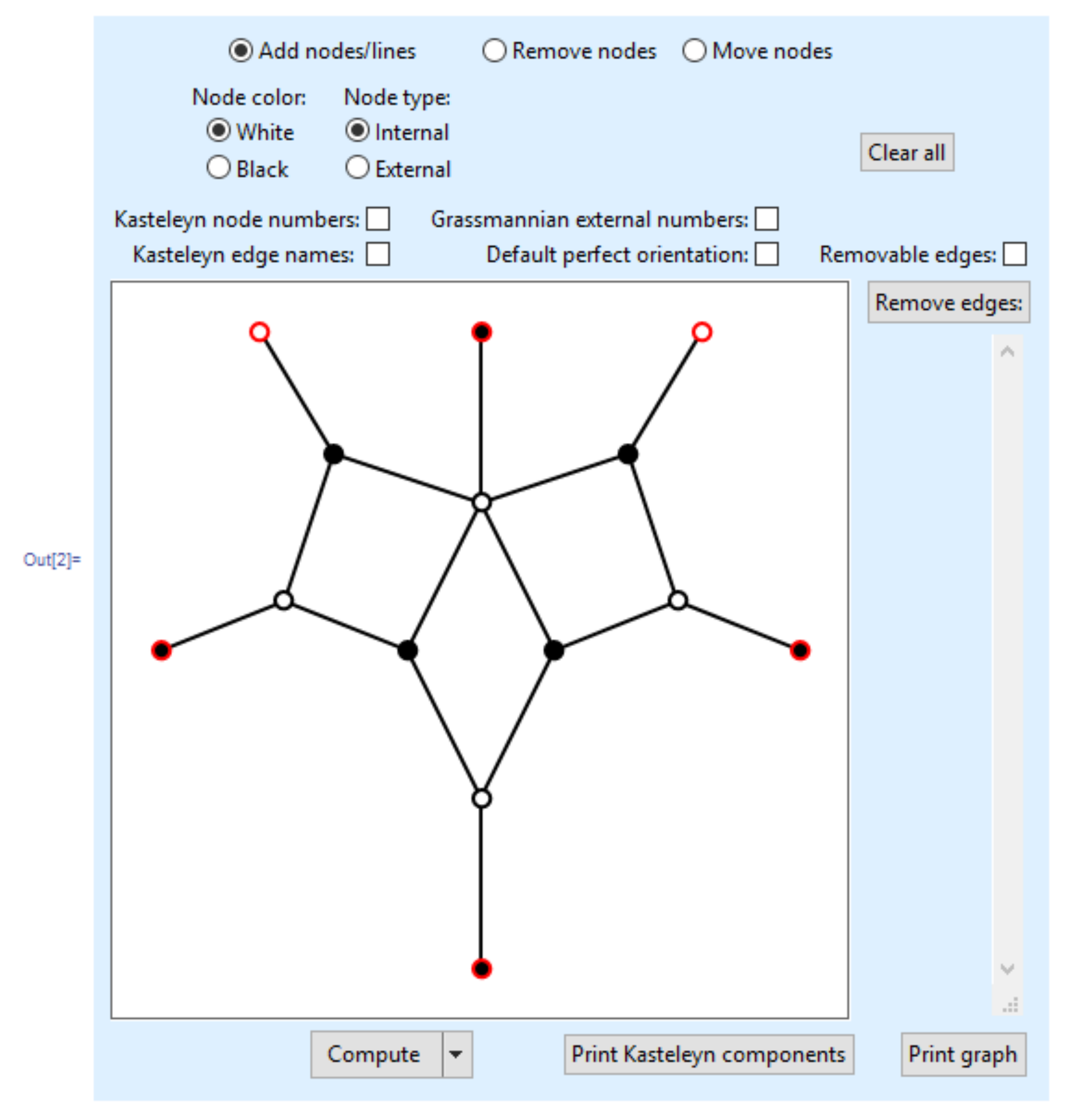}
\label{fig:drawGraphG26}
\end{center}
\end{figure}

\noindent If we now click on the button ``Print Kasteleyn components'', we obtain a print-out beneath the blue box:
\begin{flalign*}
& \pc{The Kasteleyn components are:} & \\
& \quad \pc{\{"top-left","top-right","bottom-left","bottom-right"\}} &  \\
& \pc{\{\{\{Z[1],Z[2],0,0\},\{0,Z[3],Z[4],0\},\{0,0,Z[5],Z[6]\},\{Z[7],Z[8],Z[9],Z[10]\}\},} & \\
& \pc{\{\{Z[11],0,0,0\},\{0,Z[12],0,0\},\{0,0,Z[13],0\},\{0,0,0,Z[14]\}\},} & \\
& \pc{\{\{Z[15],0,0,0\},\{0,0,0,Z[16]\}\},} & \\
& \pc{\{\{0,0,0,0\},\{0,0,0,0\}\}\}}  &
\end{flalign*}
Depending on the order in which the nodes were placed in the diagram, the actual print-out may vary---this has to do with the ordering of rows and columns in each component of the Kasteleyn matrix and is of no consequence.

As seen, the print-out contains a list with four matrices in it; these correspond to the four blocks that make up the Kasteleyn. As indicated by the print-out, the first matrix is the top-left part of the Kasteleyn in \eref{eq:kasteleynstructure}, i.e.\ the adjacency matrix between internal white nodes and internal black nodes. The second element is the top-right part, connecting internal white nodes to external black nodes. The third element is the bottom-left part. Finally, the bottom-right part is seen to only contain zeros, as required.

The \pc{Z[}$i$\pc{]} variables inside the Kasteleyn matrix are the weights given to the edges, by default labeled in numerical order. Assembling the matrix components into a single matrix (which the helper-function \pc{joinupKasteleyn} does for us), we obtain the correct Kasteleyn matrix for this diagram:
{\small
\begin{equation} \label{eq:kasteleynexample}
K = \left(
\begin{array}{cccc|cccc}
 Z[1] & Z[2] & 0 & 0 & Z[11] & 0 & 0 & 0 \\
 0 & Z[3] & Z[4] & 0 & 0 & Z[12] & 0 & 0 \\
 0 & 0 & Z[5] & Z[6] & 0 & 0 & Z[13] & 0 \\
 Z[7] & Z[8] & Z[9] & Z[10] & 0 & 0 & 0 & Z[14] \\
 \hline
 Z[15] & 0 & 0 & 0 & 0 & 0 & 0 & 0 \\
 0 & 0 & 0 & Z[16] & 0 & 0 & 0 & 0 \\
\end{array}
\right) \; .
\end{equation}
}%

Ticking the ``Kasteleyn node numbers'' box in \pc{drawGraph} helps identify which row and column corresponds to which node: the nodes are always counted beginning with white internal nodes, followed by white external nodes, black internal nodes and finally black external nodes. Ticking the ``Kasteleyn edge names'' box helps identify which \pc{Z[}$i$\pc{]} is assigned to which edge in the graph.

It is worthwhile to note what happens when we introduce additional edges between a pair of nodes\footnote{This is done by simply dragging a new line between two nodes in the interactive box which are already connected.}: the graph forms a bubble, and if we print out the new Kasteleyn, we see that one entry has turned into a sum of edges.

Finally, we point out that the function \pc{drawGraph} can actually take as input the four components of the Kasteleyn, and draw them for us in the interactive box; if we were to evaluate
\begin{flalign*}
& \quad \quad \quad \incell{3} \pc{drawGraph[topleft,topright,bottomleft,bottomright]} & 
\end{flalign*}
where \pc{topleft,topright,bottomleft,bottomright} are the four matrices given in the print-out above, we would obtain the same diagram as shown above. This feature is particularly useful when wanting to perform graphical manipulations on already-created Kasteleyns.

\subsubsection[\N{1} BFTs: Specifying a Graph Embedding]{$\boldsymbol{\mathcal{N}=1}$ BFTs: Specifying a Graph Embedding} 
\label{sec:bftembedding}

\begin{figure}[hbt]
\begin{center}
\includegraphics[scale=0.8]{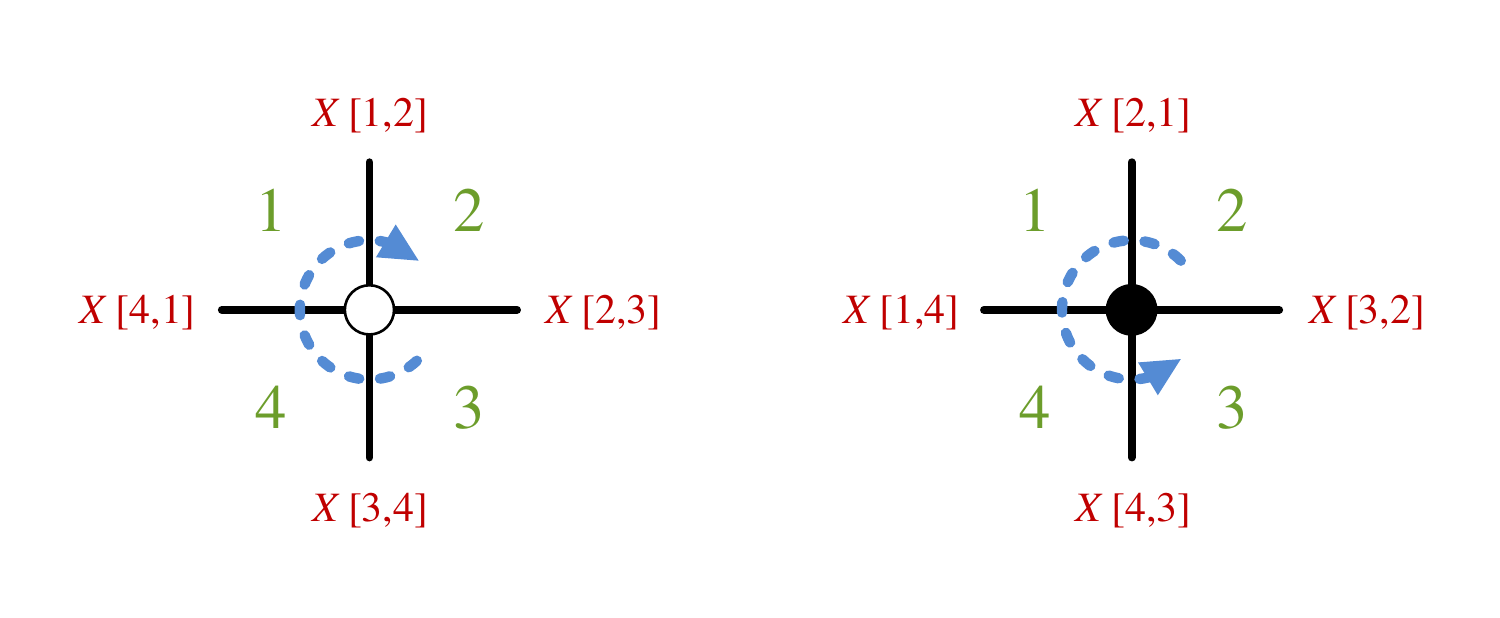}
\vspace{-0.75cm}
\caption{Edges are labeled clockwise around white nodes and counter-clockwise around black nodes, using the face-numbers as labels. Here faces are labeled in green and edges in red.}
\label{fig:edgelabeling}
\end{center}
\end{figure}

The function \pc{drawGraph} treats the diagrams as embedding-independent, i.e.\ as only depending on the nodes and their connectivity. For the purposes of \N{1} BFTs this will not be sufficient, as BFT$_1$s define the gauge groups to be the internal faces, which depend on the embedding. The information on the embedding can easily be incorporated into the Kasteleyn through the labeling of the indices of the edges. In BFT$_1$s, each face is assigned a label $i$, where $i$ is a positive integer.\footnote{The requirement that $i>0$ is purely for the stability of \pc{bipartiteSUSY}.} Edges are then labeled as \pc{X[}$i$\pc{,}$j$\pc{]}, where $i$ and $j$ are face labels, and are labeled clockwise around white nodes and counter-clockwise around black nodes, as shown in \fref{fig:edgelabeling}. We note that the face labels are not required to be \textit{consecutive} integers; aside from requiring $i>0$, they are completely free.

When dealing with BFT$_1$s it is still possible to use \pc{drawGraph} to some advantage: the diagram may be drawn interactively to generate the Kasteleyn matrix, from which it is a simple exercise to replace the \pc{Z[}$i$\pc{]} edge variables with variables consistent with a user-chosen labeling of faces. It is then possible to use various functions in the package, documented in \sref{sec:functionlist}, to check the consistency of the labeling of the BFT.

\section{Kasteleyn Matrix Format in \pc{bipartiteSUSY}}
\label{sec:kasteleynrules}

Nearly all functions in \pc{bipartiteSUSY} only require the Kasteleyn matrix as the starting point, from which it is possible to compute a large range of complicated properties and results. The format in which the Kasteleyn matrix is specified in the package must be as follows:
\begin{itemize}
\item The matrix is broken up into the four constituent matrices as shown in \eref{eq:kasteleynstructure}. In the documentation in \sref{sec:functionlist} they are usually referred to as \pc{topleft}, \pc{topright}, \pc{bottomleft} and \pc{bottomright} due to their location in the Kasteleyn.
\item Rows represent white nodes and columns represent black nodes.
\item In the absence of external white nodes, the bottom-left part has zero rows, and is hence equal to the empty matrix \pc{\{\}}. In the absence of external black nodes, the top-right part has zero columns but as many rows as the top-left part; hence, the top-right part is equal to \pc{\{\{\},\{\},\ldots,\{\}\}}, where there should be as many empty brackets as there are rows in the top-left part of the Kasteleyn.
\item In the case of \N{4} scattering amplitudes, edges in the Kasteleyn should ideally be labeled numerically (but not necessarily sequentially) using square brackets, e.g.\ \pc{Z[100]} or \pc{edge[100]}.
\item In the case of \N{1} BFTs, even though it is only BFT$_1$s that strictly speaking require an embedding, it is often useful to describe BFT$_2$s in terms of an embedding as well \cite{Franco:2012wv}. For this reason, the use of the \pc{bipartiteSUSY} package for BFTs is designed for edges labeled as described in \fref{fig:edgelabeling}.\footnote{For the majority of functions, labeling edges in BFT$_2$s in a way that does not adhere to this rule but rather takes the more lax labeling of scattering amplitudes will not cause problems. However, for reasons of stability and consistency, it is advisable to use the two-index labels of BFTs when dealing with BFT$_2$s.}
\item External nodes should have either one or zero edges connected to them.
\end{itemize}

\section{Functions} 
\label{sec:functionlist}

We shall now describe a complete list of all functionality in the \pc{bipartiteSUSY} package. All function names begin with lowercase and proceed with camelCase. The nomenclature used for the functions reflect their use and type:\footnote{It is of course not necessary to know the nomenclature to use the functions, but it is helpful in order to understand their classification, and it serves as a useful mnemonic for recalling the function names.}
\begin{itemize}
\item Functions that take as input the Kasteleyn and return some basic property of the graph which can be immediately read off from the Kasteleyn, such as whether the diagram has bubbles (i.e.\ multiple edges between a pair of nodes), or duplicate edge names, or what the sources are in a perfect orientation, have the format \pc{get<}PropertyName\pc{>}. E.g.\ \pc{getDuplicateEdges} or \pc{getSourceNodes}.
\item Functions that end in \pc{Q} return \pc{True} or \pc{False}. E.g.\ \pc{planarityQ} or \pc{reducibilityQ}.
\item Functions that simply transform the information described by the input into a different format, such as turning the Kasteleyn matrix into a bipartite graph, or an adjacency matrix that can be fed into Mathematica's graphical utilities, have the format \pc{turnInto<}ObjectName\pc{>}. E.g.\ \pc{turnIntoGraph} or \pc{turnIntoAdjacencyMatrix}.
\item Functions that take as input the Kasteleyn, compute some derived property or quantity, and return its value have the format \pc{<}computedObjectName\pc{>}. Most functions in the package are of this type. E.g.\ \pc{stratificationBoundaries}, \pc{grassmannianMatrix}, \pc{reductionGraph} or \pc{moduliSpaceBFT}.
\end{itemize}

Additionally, in what follows we distinguish those functions which are intended for general use on bipartite diagrams from those specifically intended for use in \N{4} SYM scattering amplitudes or for use in \N{1} BFTs. This distinction is made through a color-coding of the bullet points listing the functions:
\vspace{-6pt}
\begin{itemize}
\item[{\color[RGB]{0,0,0} $\bullet$}] black bullet points are for general use, \vspace{-6pt}
\item[{\color[RGB]{255,51,51} $\bullet$}] red bullet points are specifically intended for scattering amplitudes, \vspace{-6pt}
\item[{\color[RGB]{180,0,180} $\bullet$}] purple points are specifically intended for BFTs.
\end{itemize}

\subsection{Basic Functions for the Kasteleyn} 
\label{sec:basickasteleynfunctions}

Here we present a list of low-level functions that handle basic information in the Kasteleyn.

\point \func{getDuplicateEdges}{\kast}: This function takes the four components of the Kasteleyn and returns a list of edge names that are duplicates. The user should rename these, since edges must have distinct names.

\pointbft \func{getEdgesBFTformQ}{\kast}: Checks whether the edges in the Kasteleyn components have the generic form \pc{\_[\_Integer,\_Integer]}, e.g.\ \pc{X[1,2]} or \pc{edgename[1,1]}. Returns \pc{True} or \pc{False}.

\pointbft \func{getKasteleynConsistencyViolation}{\kast}: Checks whether the index structure of the Kasteleyn for the BFT was done as described in \fref{fig:edgelabeling}, i.e.\ with cyclic index labels around white and black nodes, where the second index of each edge will appear as the first index of a different edge and vice-versa. The functions returns a list of two elements, where the first element is a list of row-numbers in the Kasteleyn in which there is a problem with the index structure, and the second element is a list of column-numbers with incorrect index structure. E.g.\ the function returning \pc{\{\{2\},\{1,4\}\}} would mean there is a problem in the second row and first and fourth columns in the Kasteleyn matrix.

\point \func{getKasteleynCheckQ}{\kast,\varbar{BFTgraph}}: Checks the Kasteleyn for any of the problems that might be raised by the functions \pc{getDuplicateEdges}, \pc{getEdgesBFTformQ} or \pc{getKasteleynConsistencyViolation}. Since the latter two are only necessary for BFTs, this function requires the user to specify whether it is being used in a BFT context: \var{BFTgraph} should be either \pc{True} (in the case of BFTs) or \pc{False} (in the case of scattering amplitudes). The function returns \pc{True} or \pc{False}. Additionally, if it returns \pc{False} on a given Kasteleyn, a print-out lists the issues with the Kasteleyn.

\point \func{joinupKasteleyn}{\kast}: Returns the Kasteleyn matrix as a single matrix, by simply joining up its four components \var{topleft}, \var{topright}, \var{bottomleft} and \var{bottomright}.

\pointbft \func{getNumberFaces}{\kast}: Returns the total number of faces in the graph, as given by the indices of the edges in the Kasteleyn components.

\pointbft \func{getNumberExternalFaces}{\kast}: Returns the total number of external faces, i.e.\ those faces touching an external boundary, as given by the indices of the edges in the Kasteleyn components.

\pointbft \func{getNumberInternalFaces}{\kast}: Returns the total number of internal faces, i.e.\ those faces not touching an external boundary, as given by the indices of the edges in the Kasteleyn components.

\pointbft \func{getFaceLabels}{\kast}: Returns a list of all numeric face labels appearing in the Kasteleyn, as given by the indices of the edges in the Kasteleyn components.

\pointbft \func{getExternalFaceLabels}{\kast}: Returns a list of all numeric \textit{external} face labels appearing in the Kasteleyn, as given by the indices of the edges in the Kasteleyn components.

\pointbft \func{getInternalFaceLabels}{\kast}: Returns a list of all numeric \textit{internal} face labels appearing in the Kasteleyn, as given by the indices of the edges in the Kasteleyn components.

\subsection{Graphical Representations and Manipulations} 
\label{sec:graphingfunctions}

Here we present functions that transform the Kasteleyn into a form amenable to Mathematica's inbuilt graph-functionality.

\point \func{turnIntoGraph}{\kast,\varbardef{edgenames}{True}}: Returns a Mathematica graph\footnote{Mathematica has rather complex algorithms for determining how to best embed the graph when visualizing it; \pc{turnIntoGraph} simply uses Mathematica's default choices.} illustrating the bipartite graph described by the Kasteleyn. The final, optional, argument of the function determines whether the returned graph should have the edge labels printed half-way across each edge; omitting the argument prints out the edge labels, whereas including the fifth argument with the value \pc{False} will avoid printing out the labels. The node numbers are always displayed in black.

\point \func{drawGraph}{\varbardef{topleft}{\{\}},\varbardef{topright}{\{\}},\varbardef{bottomleft}{\{\}},\varbardef{bottomright}{\{\}}}: Returns the interactive graph-drawing utility described in \sref{sec:drawgraph}. This function can be evaluated with no arguments, \pc{drawGraph[]}, or with the four components of the Kasteleyn, \pc{drawGraph[topleft,topright,bottomleft,bottomright]}. Choosing the latter option will produce the graph-drawing utility box with the graph already drawn.

\point \func{turnIntoAdjacencyMatrix}{\kast,\newline \phantom{444444}\varbardef{edgenames}{True}}: Returns the standard adjacency matrix of the graph, i.e.\ a square matrix where rows and columns each represent both white and black nodes in the graph. Each entry in the matrix counts the number of edges between the nodes given by the row and column of the matrix entry. Because edges in the graph are not directed, the matrix will be symmetric. Applying the Mathematica in-built function \pc{AdjacencyGraph} to the adjacency matrix produced by \pc{turnIntoAdjacencyMatrix} will yield a standard, undecorated Mathematica graph (which is very useful for checking graph isomorphisms with the function \pc{IsomorphicGraphQ}).

\point \func{turnIntoWeightedAdjacencyMatrix}{\kast}: Returns the standard weighted adjacency matrix of the graph, i.e.\ a square matrix where rows and columns each represent nodes in the graph, as for \pc{turnIntoAdjacencyMatrix}, and each entry is given by the sum of edges between the nodes specified by the row and column. As for \pc{turnIntoAdjacencyMatrix}, this matrix is symmetric.

\point \func{turnIntoOrientedAdjacencyMatrix}{\kast,\newline \phantom{44444} \varbardef{referencematching}{Null},\varbardef{withedgeweights}{False}}: Returns the adjacency matrix for the graph endowed with a perfect orientation\footnote{For an introduction to perfect orientations, and their relation to perfect matchings, see e.g.\ \cite{Franco:2013nwa}.}, whose edges are \textit{directed}. As for \pc{turnIntoAdjacencyMatrix}, the adjacency matrix is square, but its orientation means that it is no longer symmetric. The function will choose a default perfect orientation with the lowest possible number of loops. It is possible to specify the perfect orientation with the optional argument \var{referencematching}, by setting it to be the product of edges of a chosen perfect matching, which in turn specifies the perfect orientation. The perfect matchings can be found with the function \pc{perfectMatchings} introduced below. The optional argument \var{withedgeweights} will yield a weighted oriented adjacency matrix when set to \pc{True}.

\subsection{Core Functions on Bipartite Graphs} 
\label{sec:generalbipartitafunctions}

Here we will list functions of general utility on bipartite graphs. Many of these form the backbone of the powerful combinatorial machinery developed for theories described by bipartite graphs.

\point \func{perfectMatchings}{\kast}: Returns a list of perfect matchings of the diagram, where each perfect matching is expressed as the product of edges used by the perfect matching. This is a key combinatorial object of bipartite graphs; the list of all perfect matchings contains the full information of the graph repackaged in a format that allows for very powerful machinery \cite{Franco:2013nwa,Franco:2014nca,Franco:2015rma}.

\point \func{lowNumberLoopsPMpos}{\kast}: Perfect matchings are one-to-one with perfect orientations. It is often desirable to deal with perfect orientations of diagrams that contain a low number of closed loops. \pc{lowNumberLoopsPMpos} returns a number indicating a position in the list returned by \pc{perfectMatchings}, containing a perfect matching whose perfect orientation has the lowest possible number of loops in the diagram. 
\begin{itemize}
\item[] {\small \textit{Example.} Let us illustrate this function with a simple example, given by the Kasteleyn
\begin{flalign*}
\incell{2} & \pc{topleft=\{\{Z[1],Z[2],Z[3]\},\{Z[4],Z[5],0\},\{0,Z[6],Z[7]\}\};} \\
& \pc{topright=\{\{0,0\},\{0,Z[8]\},\{Z[9],0\}\};} \\
& \pc{bottomleft=\{\{0,0,Z[10]\},\{Z[11],0,0\}\};} \\
& \pc{bottomright=\{\{0,0\},\{0,0\}\};} 
\end{flalign*}
We shall now show how \pc{lowNumberLoopsPMpos} works:
\begin{flalign*}
\incell{6} & \pc{matchings=perfectMatchings[topleft,topright,bottomleft,} \\
& \quad \; \; \pc{bottomright];} \\
& \pc{pos=lowNumberLoopsPMpos[topleft,topright,bottomleft,bottomright];} \\
\incell{8} & \pc{pos} \\
& \pc{matchings[[pos]]} \\
\outcell{8} & \pc{4} \\
\outcell{9} & \pc{Z[1]Z[6]Z[8]Z[10]}
\end{flalign*}
Which tells us that the fourth perfect matching, i.e.\ \pc{Z[1]Z[6]Z[8]Z[10]}, specifies a perfect orientation with the lowest possible number of loops. Using \pc{AdjacencyGraph} on \pc{turnIntoOrientedAdjacencyMatrix} it is easy to check that there are no loops in this perfect orientation.\footnote{It may be necessary to use Mathematica's \pc{GraphLayout} option to change the embedding to a more suitable one.} If there are other perfect orientations with the same number of loops, \pc{lowNumberLoopsPMpos} returns the first one.
}%
\end{itemize}

\point \func{matchingPolytope}{\kast}: Returns a matrix whose rows correspond to edges in the graph and whose columns correspond to perfect matchings. The perfect matchings are ordered according to the output of \pc{perfectMatchings}. If an edge participates in a given perfect matching, the corresponding matrix entry is \pc{1}, otherwise it is \pc{0}. Viewing the columns of this matrix as coordinates in a $N_E$-dimensional space, where $N_E$ is the number of edges, this matrix describes a set of points whose convex hull is known as the \textit{matching polytope}. The dimension of the matching polytope is usually much smaller than the dimension in which it is embedded using this parametrization. The face lattice of the matching polytope gives the full set of valid edge removals of the graph \cite{2007arXiv0706.2501P}, and is the basis of modern graph-stratification techniques used to compute the singularity structure of on-shell diagrams \cite{Franco:2013nwa,Franco:2015rma,Bourjaily:2016mnp}. The matching polytope of a BFT equals the toric diagram for the master space of the BFT.

\point \func{matchingPolytopeBoundaries}{\kast}: Obtains all boundaries of all dimensionalities of the matching polytope, i.e.\ computes its face lattice. Since each boundary of the matching polytope corresponds to the matching polytope of a sub-graph of the original graph \cite{2007arXiv0706.2501P}, each boundary is denoted with the list of edges that are present in that sub-graph. \pc{matchingPolytopeBoundaries} returns a list where each element contains all boundaries of a given codimension, starting from codimension 0 and going all the way down to dimension 0.
\begin{itemize}
\item[] {\small \textit{Example.} We shall obtain all boundaries of the matching polytope for a particularly simply example: the square box (whose Kasteleyn can be quickly obtained using \pc{drawGraph}).
\begin{flalign*}
\incell{2} & \pc{allboundaries=matchingPolytopeBoundaries[} \\
& \quad \; \; \pc{\{\{Z[1],Z[2]\},\{Z[3],Z[4]\}\},\{\{Z[5],0\},\{0,Z[6]\}\},} \\
& \quad \; \; \pc{\{\{Z[7],0\},\{0,Z[8]\}\},\{\{0,0\},\{0,0\}\}];} \\
\incell{3} & \pc{allboundaries[[1]]} \\
\outcell{3} & \pc{\{\{Z[1],Z[2],Z[3],Z[4],Z[5],Z[6],Z[7],Z[8]\}\}}\\
\incell{4} & \pc{allboundaries[[-1]]} \\
\outcell{4} & \pc{\{\{Z[5],Z[6],Z[7],Z[8]\},\{Z[4],Z[5],Z[7]\},\{Z[3],Z[5],Z[8]\},} \\
& \; \; \pc{\{Z[2],Z[6],Z[7]\},\{Z[2],Z[3]\},\{Z[1],Z[6],Z[8]\},\{Z[1],Z[4]\}\}}
\end{flalign*}
Here we see that the first element contains only one boundary, which is the list of all edges: this reflects the fact that there is a single top-dimensional object in the list of boundaries, which is the full matching polytope itself. The last element in \pc{allboundaries}, on the other hand, contains all zero-dimensional boundaries, which are seen to be simply the edges in the perfect matchings (as can also be confirmed using \pc{perfectMatchings}).
}%
\end{itemize}

\point \func{matchingPolytopeBoundariesGraph}{\kast}: Returns the graph of the face lattice of the matching polytope, i.e.\ its stratification graph. The function returns a Mathematica graph object; each node is an element of the face lattice, i.e.\ a boundary of the matching polytope of some dimension; arrows indicate the codimension-1 sub-boundaries of each boundary. To view the graph in the canonical layered way, where each vertical layer corresponds to a given dimension (with the top-dimensional element on top), it may be necessary to specify the embedding by hand, by evaluating \pc{Graph[graph, GraphLayout$\mathtt{\to}$"{\color[RGB]{102,102,102} LayeredDigraphEmbedding}"]}, where \pc{graph} is a variable that stores the object returned by \pc{matchingPolytopeBoundariesGraph}.
\begin{itemize}
\item[] {\small \textit{Example.} For the square box diagram used to exemplify \pc{matchingPolytopeBoundaries}, we obtain the diagram (after specifying the \pc{GraphLayout} as described)
\begin{figure}[hbt]
\begin{center}
\includegraphics[width=\textwidth]{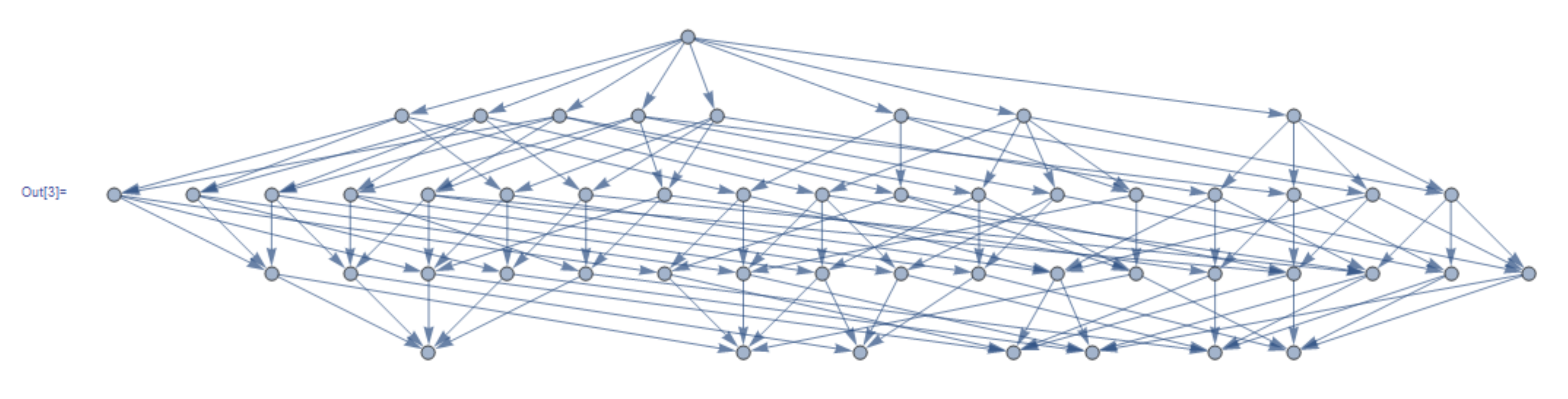}
\label{fig:sqbfacelattice}
\end{center}
\end{figure}
}%
\end{itemize}

\point \func{squareMove}{\kast,\varbar{fournodesorfacenum},\newline \phantom{444444}\varbardef{BFTgraph}{False}}: Returns a list containing the four Kasteleyn components, in the schematic form \pc{\{topleft,} \pc{topright,bottomleft,bottomright\}}, after having performed a square move on a four-sided cycle. The cycle is specified in the fifth entry of the input, as a list of four node numbers participating in the cycle. In the case of BFTs it can alternatively be specified by the face number on which to perform the square move, as long as the edges have been labeled according to \fref{fig:edgelabeling} and the optional entry \var{BFTgraph} is set to \pc{True}.
\begin{itemize}
\item[]
{\small \textit{Example.} Let us illustrate how this works on a simple example, given by the Kasteleyn 
\begin{flalign*}
\incell{1} & \pc{topleft=\{\{X[3,1],X[1,2],X[2,3]\},\{X[1,4],X[5,1],0\},\{0,X[2,5],X[6,2]\}\};} \\
& \pc{topright=\{\{0,0\},\{0,X[4,5]\},\{X[5,6],0\}\};} \\
& \pc{bottomleft=\{\{0,0,X[3,6]\},\{X[4,3],0,0\}\};} \\
& \pc{bottomright=\{\{0,0\},\{0,0\}\};}
\end{flalign*}
The graph can be easily viewed by evaluating \pc{turnIntoGraph[topleft,topright,} \pc{bottomleft,bottomright]} and is displayed below.
\begin{figure}[hbt]
\begin{center}
\includegraphics[scale=0.2]{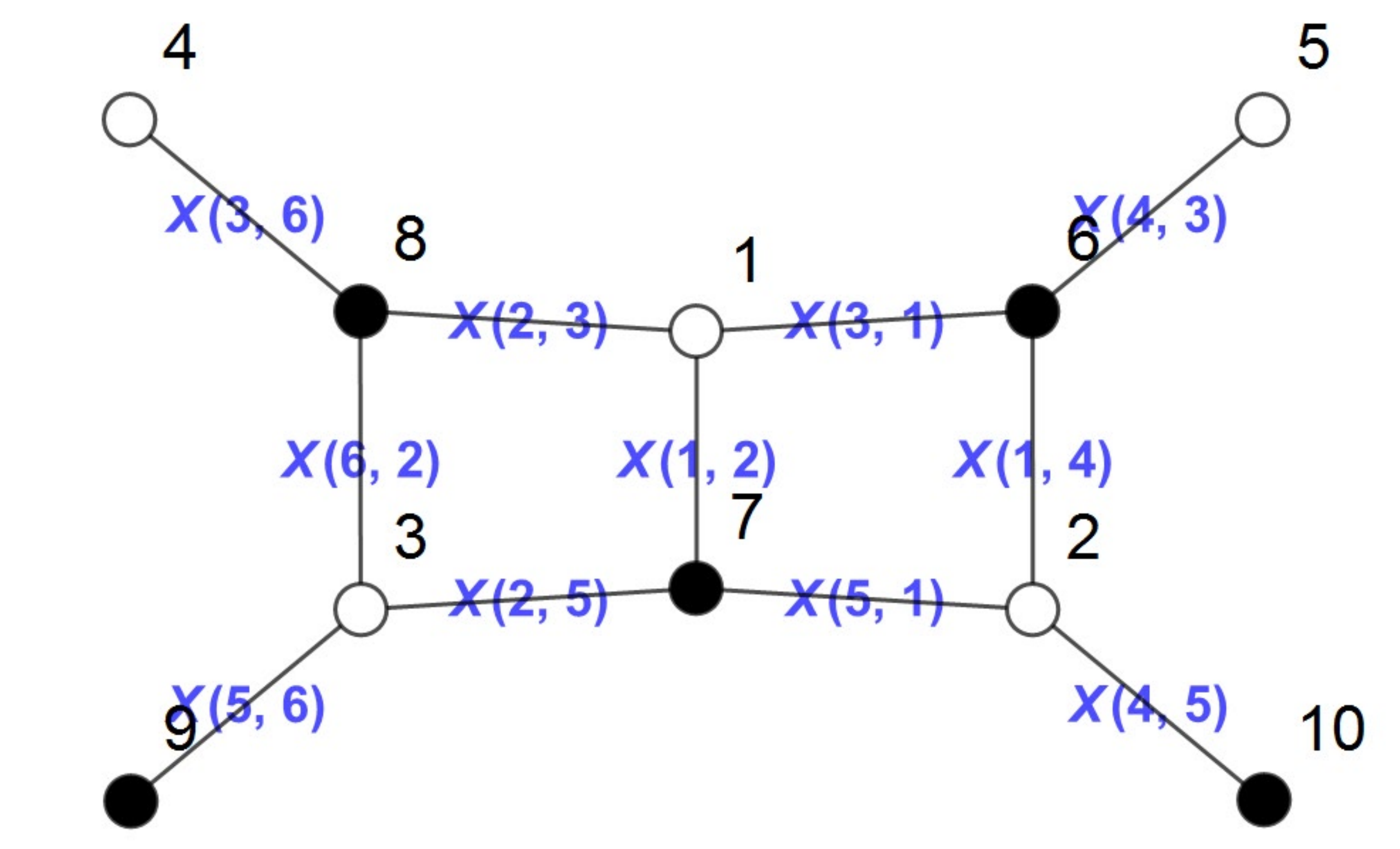}
\label{fig:doublesquarebox}
\end{center}
\end{figure}

We shall now perform a square move on the four-edge cycle given by nodes \pc{\{1,6,2,7\}} as shown in black in the figure:
\begin{flalign*}
\incell{2} & \pc{squareMove[topleft,topright,bottomleft,bottomright,\{1,2,6,7\}]} \\
\outcell{2} & \pc{\{\{\{0,0,X[2,3],XX[9],0\},\{0,0,0,0,XX[10]\},\{0,X[2,5],X[6,2],0,0\},} \\
& \pc{\{XX[7],0,0,X[3,1],X[1,2]\},\{0,XX[8],0,X[1,4],X[5,1]\}\},\{\{0,0\},}\\
& \pc{\{0,X[4,5]\},\{X[5,6],0\},\{0,0\},\{0,0\}\},\{\{0,0,X[3,6],0,0\},} \\
& \pc{\{X[4,3],0,0,0,0\}\},\{\{0,0\},\{0,0\}\}\}}
\end{flalign*}
where we note that the nodes are not required to be specified in any determined order. The diagram for the returned Kasteleyn is shown below, where Mathematica has by default mirrored left and right in the embedding, with respect to the previous figure.
\begin{figure}[hbt]
\begin{center}
\includegraphics[scale=0.3]{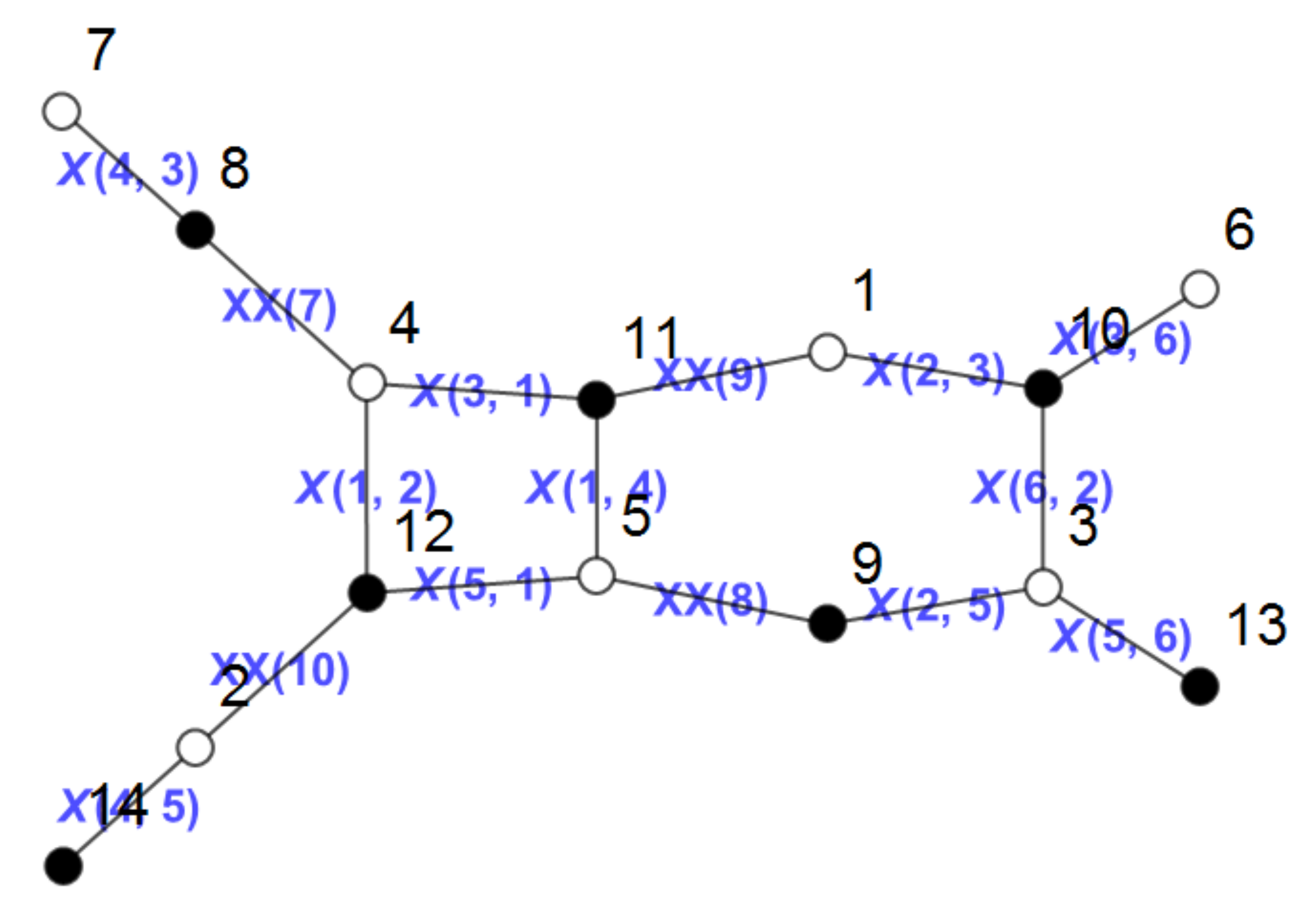}
\label{fig:doublesquareboxsqmv}
\end{center}
\end{figure}

If we were interested in BFTs and wanted to preserve the BFT labeling of edges given by \fref{fig:edgelabeling}, we would simply need to evaluate 
\begin{flalign*}
\incell{2} & \pc{squareMove[topleft,topright,bottomleft,bottomright,\{1,2,6,7\},True]} 
\end{flalign*}
which would produce an equivalent output to 
\begin{flalign*}
\incell{2} & \pc{squareMove[topleft,topright,bottomleft,bottomright,1,True]} \quad \quad \quad \quad  
\end{flalign*}
}%
\end{itemize}

\point \func{collapseBivalentNodes}{\kast}: Returns a list containing the four Kasteleyn components, in the schematic form \pc{\{topleft,topright,} \pc{bottomleft,bottomright\}}, after having collapsed bivalent nodes in the diagram.

\point \func{bubblesQ}{\kast,\varbardef{BFTgraph}{False},\varbardef{gauging}{2}}: Establishes whether the diagram has bubbles, which for scattering amplitudes and BFT$_2$s are defined as multiple edges between the same two nodes, and for BFT$_1$s are defined as faces with only two edges. The default for the function is to treat the diagram as the former; to deal with BFT$_1$s it is necessary to specify the two optional arguments as \pc{True} and \pc{1}. The function returns \pc{True} or \pc{False}.

\point \func{removeBubbles}{\kast,\varbardef{BFTgraph}{False},\newline \phantom{444444}\varbardef{gauging}{2}}: Returns a list containing the four Kasteleyn components, in the schematic form \pc{\{topleft,topright,bottomleft,bottomright\}}, after having removed bubbles. As for the previous function, to deal with BFT$_1$s it is necessary to specify the two optional arguments as \pc{True} and \pc{1}.

\point \func{simplifyGraph}{\kast}: Returns a list containing the four Kasteleyn components, in the schematic form \pc{\{topleft, topright,} \pc{bottomleft,bottomright\}}, after iteratively removing bubbles and collapsing bivalent nodes until it is no longer possible to do so. Here any pair of edges between two given nodes are seen as forming a bubble, i.e.\ the graph is treated as a scattering graph, or equivalently a BFT$_2$. This function will not necessarily achieve a maximal reduction, but is computationally very fast and can be complemented with \pc{reductionGraph} or \pc{reductionGraphBFT}, introduced in \sref{sec:scatteringinfofunctions} and \sref{sec:bftfunctions}, respectively.

\point \func{consistentEdgeRemoval}{\kast,\varbar{edgelist}}: Sometimes removing a set of edges from a graph will yield a diagram containing edges that do not participate in any perfect matching. These edges are superfluous and should also be removed for a consistent edge removal \cite{2007arXiv0706.2501P}. \pc{consistentEdgeRemoval} requires as input the four Kasteleyn components and the list of edges \var{edgelist} to be removed; the function then returns the full list of edges that form a consistent edge removal.

\point \func{survivingPerfectMatchings}{\kast,\varbar{edgelist}}: Removing edges will kill all perfect matchings that use those edges. This function returns a list of numbers, which specify the positions of the perfect matchings given by \pc{perfectMatchings} which survive after removing the edges specified by \var{edgelist}.
\begin{itemize}
\item[]
{\small \textit{Example.} Let us see how this works in the simple example of the square box, where we remove the edges \pc{Z[1]} and \pc{Z[2]}:
\begin{flalign*}
\incell{2} & \pc{survivingPerfectMatchings[\{\{Z[1],Z[2]\},\{Z[3],Z[4]\}\},\{\{Z[5],0\},} \\
& \; \; \pc{\{0,Z[6]\}\},\{\{Z[7],0\},\{0,Z[8]\}\},\{\{0,0\},\{0,0\}\},\{Z[1],Z[2]\}]} \\
\outcell{2} & \pc{\{3,5,7\}}
\end{flalign*}
which tells us that after removing \pc{Z[1]} and \pc{Z[2]}, the only perfect matchings that survive are found at positions 3, 5 and 7 in the list of perfect matchings given by \pc{perfectMatchings}. It is easy to check that these perfect matchings are the only ones that do not contain \pc{Z[1]} or \pc{Z[2]}.
}%
\end{itemize}

\point \func{loopVariablesBasis}{\kast,\newline \phantom{444444}\varbardef{standardfacevariables}{False}}: Loop variables form a basis of paths. This function returns a list of loop variables, where each path is expressed as a product of edges. If the path traverses an edge from a white node to a black node the edge appears in the numerator of the expression, otherwise it appears in the denominator. The paths given by \pc{loopVariablesBasis} are by default divided into two lists: those which are ``internal'', i.e.\ form independent closed loops, and those which are ``external'', i.e.\ use external edges. General loop variables do not require an embedding and can be computed without a BFT-labeling of edges, and indeed are useful in scattering amplitudes. Canonical face variables are a specific choice of loop variables that use the faces of the graph as a basis, as described in \cite{Franco:2012wv}. For BFTs the first list represents those symmetries which are gauged, and the second list represents those symmetries which are global symmetries. For BFT$_1$s, non-trivial loops around higher-genus surfaces are not gauged; for this reason, when setting the final optional entry \var{standardfacevariables} to \pc{True}, the non-trivial loops around the higher-genus surface will appear in the second list, despite being ``internal'' loops. \pc{loopVariablesBasis} will always return the minimal number of paths.
\begin{itemize}
\item[] {\small \textit{Example.} We shall present an example which, if embedded on a higher-genus surface and edges are labeled according to \fref{fig:edgelabeling}, has the Kasteleyn components
\begin{flalign*}
\incell{1} & \pc{topleft=\{\{X[2,1]+X[3,4],X[1,3]+X[4,2]\},\{Y[1,3]+Y[4,2],Y[2,1]\}\};} \\
& \pc{topright=\{\{0\},\{Y[3,4]\}\};} \\
& \pc{bottomleft=\{\{0,Z[3,4]\}\};} \\
& \pc{bottomright=\{\{0\}\};}
\end{flalign*}
This example transparently illustrates the difference between setting \var{standardfacevariables} to \pc{True} or \pc{False}:
\begin{flalign*}
\incell{2} & \pc{loopVariablesBasis[topleft,topright,bottomleft,bottomright]} \\
\outcell{2} & \pc{\{\{}\frac{\pc{X[2,1]Y[2,1]}}{\pc{X[1,3]Y[1,3]}},\frac{\pc{Y[4,2]}}{\pc{Y[1,3]}},\frac{\pc{X[4,2]}}{\pc{X[1,3]}},\frac{\pc{X[3,4]}}{\pc{X[2,1]}}\},\{\frac{\pc{Y[2,1]}}{\pc{Y[3,4]Z[3,4]}}\}\} \\
\incell{3} & \pc{loopVariablesBasis[topleft,topright,bottomleft,bottomright,True]} \\
\outcell{3} & \pc{\{\{}\frac{\pc{X[2,1]Y[2,1]}}{\pc{X[1,3]Y[1,3]}},\frac{\pc{X[4,2]Y[4,2]}}{\pc{X[2,1]Y[2,1]}}\},\{\frac{\pc{X[1,3]Y[1,3]}}{\pc{X[3,4]Y[3,4]Z[3,4]}},\frac{\pc{Y[4,2]}}{\pc{Y[1,3]}},\frac{\pc{X[3,4]}}{\pc{X[2,1]}}\}\}
\end{flalign*}
There is a considerable reshuffling of the loop d.o.f.\ in the second case. This is unimportant; what is of relevance is that both scenarios have five loop d.o.f., but the second scenario moved two d.o.f.\ from the first list to the second list. These two d.o.f.\ correspond to the two fundamental loops around the torus.
}%
\end{itemize} 

\point \func{matroidPolytope}{\kast}: Returns the matroid polytope associated to the graph, in the form of a matrix where the columns are vertex-coordinates in the polytope. Each column corresponds to the coordinate associated to a perfect matching. Since different perfect matchings may have the same coordinate in the matroid polytope, some columns will in general be identical. The perfect matchings are ordered according to the output of \pc{perfectMatchings}.

\point \func{turnIntoPolytope}{\varbar{matrix}}: Often \pc{matroidPolytope} will return duplicate columns. \pc{turnIntoPolytope} removes the duplicate columns and gives a tally for each distinct coordinate, i.e.\ the multiplicity of each coordinate. The function returns a list of two items: the first is the matrix containing only the distinct columns, and the second is a list where each element is the multiplicity of the corresponding column.
\begin{itemize}
\item[] 
{\small \textit{Example.} This example illustrates how \pc{matroidPolytope} returns a coordinate for each of the 10 perfect matchings in a specific example, and how this corresponds to a polytope with only 6 vertices. The multiplicities of each vertex are given by the second element of the list returned by \pc{turnIntoPolytope}.
\begin{flalign*}
\incell{2} & \pc{coordinates=matroidPolytope[\{\{Z[1],Z[2],Z[3]\},\{Z[4],Z[5],0\},} \\
   & \; \; \; \; \pc{\{0,Z[6],Z[7]\}\},\{\{0,0\},\{0,Z[8]\},\{Z[9],0\}\},\{\{0,0,Z[10]\},} \\
   & \; \; \; \; \pc{\{Z[11],0,0\}\},\{\{0,0\},\{0,0\}\}]} \\
\outcell{2} & \pc{\{\{0,0,0,1,0,0,1,1,0,1\},\{0,0,0,0,1,1,0,0,1,1\},} \\
 & \pc{ \{0,0,0,1,1,1,0,0,0,1\},\{0,0,0,0,0,0,1,1,1,1\}\}} \\
\incell{3} & \pc{turnIntoPolytope[coordinates]} \\
\outcell{3} & \pc{\{\{0,1,0,1,0,1\},\{0,0,1,0,1,1\},\{0,1,1,0,0,1\},\{0,0,0,1,1,1\}\},} \\
& \pc{ \{3,1,2,2,1,1\}\}}
\end{flalign*}
}%
\end{itemize}

\point \func{dimensionPolytope}{\varbar{matrix}}: Returns the dimension of the polytope whose vertex coordinates are given by the columns on the input-matrix \var{matrix}. Duplicate columns are permitted.
\begin{itemize}
\item[] 
{\small \textit{Example.} We shall borrow the same example used to illustrate \pc{turnIntoPolytope} and see that the matroid polytope is in reality 3-dimensional despite being embedded in 4 dimensions:
\begin{flalign*}
\incell{4} & \pc{dimensionPolytope[coordinates]} \quad \quad \quad \quad \quad \quad \quad \quad \quad \quad \quad \quad \quad \quad \quad \quad \\
\outcell{4} & \pc{3}
\end{flalign*}
}%
\end{itemize}

\pointbft \func{cyclicEdgeOrderings}{\varbar{edges},\varbar{firstedge}}: Labeling the edges with a cyclic structure around each node, as shown in \fref{fig:edgelabeling}, effectively specifies the embedding of the graph. It is often useful to be able to explicitly reconstruct the cyclic ordering of edges, given an unordered list of edges attached to a node. \pc{cyclicEdgeOrderings} takes as input a list of edges  \var{edges} and a ``first edge'' \var{firstedge} and returns all possible cyclically ordered lists of the edges in \var{edges}, which begin with \var{firstedge}.
\begin{itemize}
\item[] 
{\small \textit{Example.} Let us assume we are given the edges \pc{\{X[1,1],X[1,5],X[2,1],X[3,2],X[5,3],Y[1,1]\}} around a node in a diagram. This list has two possible (physically equivalent in this case) cyclic orderings (starting from \pc{X[1,5]}, which was chosen at random):
\begin{flalign*}
\incell{2} & \pc{cyclicEdgeOrderings[\{X[1,1],X[1,5],X[2,1],X[3,2],X[5,3],Y[1,1]\},}\\
 & \pc{ X[1,5]]} \\
\outcell{2} & \pc{\{\{X[1,5],X[5,3],X[3,2],X[2,1],X[1,1],Y[1,1]\},}\\ 
& \pc{ \{X[1,5],X[5,3],X[3,2],X[2,1],Y[1,1],X[1,1]\}\}}
\end{flalign*}
From this we see that there are two possible embeddings of these edges: either the edges are cyclically ordered with \pc{X[1,1]} appearing before \pc{Y[1,1]}, or with \pc{X[1,1]} appearing after \pc{Y[1,1]}.
}%
\end{itemize}

\subsection{The Map to the Grassmannian}
\label{sec:grassmannianfunctions}

Each bipartite graph can be mapped to an element of the Grassmannian $G(k,n)$ \cite{2006math09764P,2009arXiv0901.0020G,ArkaniHamed:2012nw,Franco:2013nwa,Franco:2015rma}. Here we shall present a list of functions that perform this operation, with and without the canonical signs in the boundary measurement. We shall also present functions that obtain the \pl coordinates, both in gauge-fixed and gauge-unfixed form, the latter being strongly superior to the former, but less known and used in the literature.

\pointscat \func{pathMatrix}{\kast,\varbardef{referencematching}{Null}}: Returns the path matrix for the on-shell diagram described by the Kasteleyn, i.e.\ a matrix of connected paths from each of the $k$ sources to each of the $n$ sinks in a perfect orientation of the graph. Each path is expressed as a product of edges, where the edge appears in the numerator of the expression if the path goes from white node to black node, and otherwise appears in the denominator. 
\begin{flushleft}
\textit{For practical purposes}, whenever obtaining the element of the Grassmannian associated to an on-shell diagram, \textit{this function is the fastest and simplest one}, and is hence the recommended one to use.
\end{flushleft}
\pc{pathMatrix} will automatically choose a perfect orientation that has the lowest possible number of loops; it is however possible to specify a perfect orientation by setting \var{referencematching} to a user-chosen perfect matching, as for \pc{turnIntoOrientedAdjacencyMatrix}. If there are loops to sum over, the function will automatically perform the geometric sum over all paths and express the loops with a factor of $(1-loop_i)$ in the denominator.

The ordering of external nodes, and hence the ordering of columns of the path matrix, is as follows: the counting begins with the external white nodes, i.e.\ sequentially from the first to the last row of \var{bottomleft}, followed by the external black nodes, i.e.\ sequentially from the first to the last column of \var{topright}.
\begin{itemize}
\item[] 
{\small \textit{Example.} We shall illustrate how to obtain the path matrix for the simplest non-planar on-shell diagram, found in $G(3,5)$ .
\begin{flalign*}
\incell{2} & \pc{pathmat=pathMatrix[\{\{Z[1],Z[2],Z[3],0\},\{Z[4],Z[5],0,Z[6]\},} \\
& \quad \; \; \pc{\{0,0,Z[7],Z[8]\}\},\{\{0\},\{0\},\{Z[9]\}\},\{\{Z[10],0,0,0\},\{0,Z[11],0,0\},} \\
& \quad \; \; \pc{\{0,0,Z[12],0\},\{0,0,0,Z[13]\}\},\{\{0\},\{0\},\{0\},\{0\}\}];} \\
& \pc{pathmat}\:\pc{//}\:\pc{MatrixForm} \\
\outcell{3} & \left(
\begin{array}{ccccc}
 \frac{\pc{Z[1]}\pc{Z[12]}}{\pc{Z[3]}\pc{Z[10]}} & \frac{\pc{Z[2]}\pc{Z[12]}}{\pc{Z[3]}\pc{Z[11]}} & \pc{1} & \pc{0} & \pc{0} \\
 \frac{\pc{Z[4]}\pc{Z[13]}}{\pc{Z[6]}\pc{Z[10]}} & \frac{\pc{Z[5]}\pc{Z[13]}}{\pc{Z[6]}\pc{Z[11]}} & \pc{0} & \pc{1} & \pc{0} \\
 \frac{\pc{Z[1]}\pc{Z[7]}}{\pc{Z[3]}\pc{Z[9]}\pc{Z[10]}}\pc{+}\frac{\pc{Z[4]}\pc{Z[8]}}{\pc{Z[6]}\pc{Z[9]}\pc{Z[10]}} & \frac{\pc{Z[2]}\pc{Z[7]}}{\pc{Z[3]}\pc{Z[9]}\pc{Z[11]}}\pc{+}\frac{\pc{Z[5]}\pc{Z[8]}}{\pc{Z[6]}\pc{Z[9]}\pc{Z[11]}} & \pc{0} & \pc{0} & \pc{1} \\
\end{array}
\right)
\end{flalign*}
The automatically chosen perfect orientation is seen to be free of loops, by noting that the denominators do not have multiple terms. If we were to instead choose the perfect orientation corresponding to perfect matching \pc{Z[2]Z[6]Z[7]Z[10]}, which is the first in the list of perfect matchings given by \pc{perfectMatchings}, we instead obtain 
\begin{flalign*}
\incell{4} & \pc{pathmat=pathMatrix[\{\{Z[1],Z[2],Z[3],0\},\{Z[4],Z[5],0,Z[6]\},} \\
& \quad \; \; \pc{\{0,0,Z[7],Z[8]\}\},\{\{0\},\{0\},\{Z[9]\}\},\{\{Z[10],0,0,0\},\{0,Z[11],0,0\},} \\
& \quad \; \; \pc{\{0,0,Z[12],0\},\{0,0,0,Z[13]\}\},\{\{0\},\{0\},\{0\},\{0\}\},Z[2]Z[6]Z[7]Z[10]];} \\
& \pc{pathmat}\:\pc{//}\:\pc{MatrixForm} \\
\outcell{3} & \hspace{-6pt} \left(
\hspace{-2pt}
\begin{array}{@{\hskip 0pt}c@{\hskip 2pt}c@{\hskip 2pt}c@{\hskip 2pt}c@{\hskip 2pt}c@{\hskip 0pt}}
 \frac{\pc{Z[1]}\pc{Z[11]}}{\pc{Z[2]}\left(\pc{1-}\frac{\pc{Z[3]}\pc{Z[5]}\pc{Z[8]}}{\pc{Z[2]}\pc{Z[6]}\pc{Z[7]}}\right)\pc{Z[10]}}\pc{+}\frac{\pc{Z[3]}\pc{Z[4]}\pc{Z[8]}\pc{Z[11]}}{\pc{Z[2]}\pc{Z[6]}\pc{Z[7]}\left(\pc{1-}\frac{\pc{Z[3]}\pc{Z[5]}\pc{Z[8]}}{\pc{Z[2]}\pc{Z[6]}\pc{Z[7]}}\right)\pc{Z[10]}}&\pc{1}&\pc{0}&\pc{0}&\frac{\pc{Z[3]}\pc{Z[9]}\pc{Z[11]}}{\pc{Z[2]}\pc{Z[7]}\left(\pc{1-}\frac{\pc{Z[3]}\pc{Z[5]}\pc{Z[8]}}{\pc{Z[2]}\pc{Z[6]}\pc{Z[7]}}\right)}\\
\frac{\pc{Z[4]}\pc{Z[8]}\pc{Z[12]}}{\pc{Z[6]}\pc{Z[7]}\left(\pc{1-}\frac{\pc{Z[3]}\pc{Z[5]}\pc{Z[8]}}{\pc{Z[2]}\pc{Z[6]}\pc{Z[7]}}\right)\pc{Z[10]}}\pc{+}\frac{\pc{Z[1]}\pc{Z[5]}\pc{Z[8]}\pc{Z[12]}}{\pc{Z[2]}\pc{Z[6]}\pc{Z[7]}\left(\pc{1-}\frac{\pc{Z[3]}\pc{Z[5]}\pc{Z[8]}}{\pc{Z[2]}\pc{Z[6]}\pc{Z[7]}}\right)\pc{Z[10]}}&\pc{0}&\pc{1}&\pc{0}&\frac{\pc{Z[9]}\pc{Z[12]}}{\pc{Z[7]}\left(\pc{1-}\frac{\pc{Z[3]}\pc{Z[5]}\pc{Z[8]}}{\pc{Z[2]}\pc{Z[6]}\pc{Z[7]}}\right)}\\
\frac{\pc{Z[4]}\pc{Z[13]}}{\pc{Z[6]}\left(\pc{1-}\frac{\pc{Z[3]}\pc{Z[5]}\pc{Z[8]}}{\pc{Z[2]}\pc{Z[6]}\pc{Z[7]}}\right)\pc{Z[10]}}\pc{+}\frac{\pc{Z[1]}\pc{Z[5]}\pc{Z[13]}}{\pc{Z[2]}\pc{Z[6]}\left(\pc{1-}\frac{\pc{Z[3]}\pc{Z[5]}\pc{Z[8]}}{\pc{Z[2]}\pc{Z[6]}\pc{Z[7]}}\right)\pc{Z[10]}}&\pc{0}&\pc{0}&\pc{1}&\frac{\pc{Z[3]}\pc{Z[5]}\pc{Z[9]}\pc{Z[13]}}{\pc{Z[2]}\pc{Z[6]}\pc{Z[7]}\left(\pc{1-}\frac{\pc{Z[3]}\pc{Z[5]}\pc{Z[8]}}{\pc{Z[2]}\pc{Z[6]}\pc{Z[7]}}\right)}\\
\end{array}
\hspace{-2pt}
\right)
\end{flalign*}
which is seen to have loops, by the presence of multiple terms in the denominator. Moreover, in this form it is transparent that there is a single loop: $\frac{Z[3]Z[5]Z[8]}{Z[2]Z[6]Z[7]}$. Finally, we stress that regardless of perfect orientation the ordering of external nodes is dictated by the rows of \var{bottomleft} and the columns of \var{topright}: we begin with the nodes attached to the edges \pc{Z[10]}, \pc{Z[11]}, \pc{Z[12]}, \pc{Z[13]} and continue with the node attached to \pc{Z[9]}.
}%
\end{itemize}

\pointscat \func{traditionalPathMatrix}{\kast,\newline \phantom{444444}\varbardef{referencematching}{Null}}: Returns the exact same matrix as \pc{pathMatrix}, albeit computed using the techniques in Appendix A of \cite{Franco:2013nwa}, which involve inverting a particular connectivity matrix. The function \pc{pathMatrix} relies on in-built graphical functionality in Mathematica, which is only available in newer versions of the program; it does not utilize the ``traditional'' techniques employed by \pc{traditionalPathMatrix}. While more reliable, insomuch as there is no reference to Mathematica's evolving graphical methods, \pc{traditionalPathMatrix} is generally \textit{much} slower than \pc{pathMatrix} (except for very small diagrams, for which there is a small speed increase); the difference is particularly dramatic in the presence of multiple loops in the perfect orientation. This function was included in the package for consistency with the tools provided in the literature, and for compatibility with older versions of Mathematica.

\pointscat \func{connectivityMatrix}{\kast,\newline \phantom{444444}\varbardef{referencematching}{Null}}: Returns a matrix containing all connected paths between all nodes in the on-shell diagram, including internal ones, in a perfect orientation of the graph. The path matrix given by \pc{pathMatrix} is simply a sub-matrix of the connectivity matrix given by this function. The connectivity matrix is a square matrix, where each row (and each column) corresponds to a different node, including both white and black nodes. The nodes are numbered in the canonical way, described under \eref{eq:kasteleynexample}. As for \pc{pathMatrix}, loops are automatically resummed into a factor of $(1-loop_i)$ in the denominator.

\pointscat \func{traditionalConnectivityMatrix}{\kast,\newline \phantom{444444}\varbardef{referencematching}{Null}}: Returns the exact same matrix as \pc{connectivityMatrix}, albeit computed using the techniques of \cite{Franco:2013nwa}, as for \pc{traditionalPathMatrix}. The same statements regarding the trade-off between speed and backwards-compatibility are true for \pc{traditionalConnectivityMatrix} as for \pc{traditionalPathMatrix}.

\pointscat \func{loopDenominator}{\kast,\newline \phantom{444444}\varbardef{referencematching}{Null},\varbardef{withsigns}{False}}: Perfect orientations, which are one-to-one with perfect matchings, may in general contain many loops. This may in turn result in a complicated-looking denominator occasionally present in the path matrix and connectivity matrix, due to the resummation of paths circling loops. It is often useful to isolate this part of the denominator; the helper function \pc{loopDenominator} does precisely this: it returns the loop part of the denominator. Moreover, the original definition of the boundary measurement, which maps on-shell diagrams to the Grassmannian, included particular signs to ensure manifest positivity of planar diagrams; setting the optional variable \var{withsigns} to \pc{True} will introduce these signs correctly.\footnote{This signs can be rather non-trivial: for multiple loops it is often necessary to also include the product of multiple loops---these terms have the sign $(-1)^L$, where $L$ is the number of multiplied loops.} As for the above functions, the user can choose a perfect orientation by setting \var{referencematching} to the corresponding perfect matching; otherwise, the program will automatically choose the perfect orientation with the lowest possible number of loops, using \pc{lowNumberLoopsPMpos}, and will hence often return \pc{1}.
\begin{itemize}
\item[] 
{\small \textit{Example.} We shall illustrate \pc{loopDenominator} for the non-planar example also used to illustrate \pc{pathMatrix}.
\begin{flalign*}
\incell{2} & \pc{loopDenominator[\{\{Z[1],Z[2],Z[3],0\},\{Z[4],Z[5],0,Z[6]\},} \\
& \quad \; \; \pc{\{0,0,Z[7],Z[8]\}\},\{\{0\},\{0\},\{Z[9]\}\},\{\{Z[10],0,0,0\},\{0,Z[11],0,0\},} \\
& \quad \; \; \pc{\{0,0,Z[12],0\},\{0,0,0,Z[13]\}\},\{\{0\},\{0\},\{0\},\{0\}\},Z[2]Z[6]Z[7]Z[10]]} \\
 & \\
 & \pc{loopDenominator[\{\{Z[1],Z[2],Z[3],0\},\{Z[4],Z[5],0,Z[6]\},} \\
& \quad \; \; \pc{\{0,0,Z[7],Z[8]\}\},\{\{0\},\{0\},\{Z[9]\}\},\{\{Z[10],0,0,0\},\{0,Z[11],0,0\},} \\
& \quad \; \; \pc{\{0,0,Z[12],0\},\{0,0,0,Z[13]\}\},\{\{0\},\{0\},\{0\},\{0\}\},Z[2]Z[6]Z[7]Z[10],} \\
 & \quad \; \; \pc{True]} \\
\outcell{2} & \pc{1-}\frac{\pc{Z[3]}\pc{Z[5]}\pc{Z[8]}}{\pc{Z[2]}\pc{Z[6]}\pc{Z[7]}} \\
\outcell{3} & \pc{1+}\frac{\pc{Z[3]}\pc{Z[5]}\pc{Z[8]}}{\pc{Z[2]}\pc{Z[6]}\pc{Z[7]}}
\end{flalign*}
}%
\end{itemize}

\pointscat \func{grassmannianMatrix}{\kast,\newline \phantom{444444}\varbardef{referencematching}{Null},\varbardef{externalordering}{Null},\varbardef{boundarylist}{Null},\newline \phantom{444444}\varbardef{boundarycutreplacements}{Null}}: Returns the matrix of the Grassmannian produced by the original definition of the boundary measurement, which includes the signs introduced for the manifest positivity of planar diagrams \cite{2006math09764P}, but also includes additional signs for non-planar diagrams \cite{2009arXiv0901.0020G,Franco:2013nwa}.\footnote{Correct signs are necessary to have a consistent map between \pl coordinates and perfect matchings. There are currently only two known options: either no signs at all, which yields the path matrix as given by \pc{pathMatrix}, or \textit{all} the signs of the boundary measurement, as given by \pc{grassmannianMatrix}.} Since the signs required for diagrams which cannot be embedded on  genus zero, as described in \cite{Franco:2015rma}, are much more challenging to implement, this function will return \pc{Null} for these cases, along with a print-out: \pc{The} \pc{diagram} \pc{cannot} \pc{be} \pc{embedded} \pc{on} \pc{genus} \pc{zero.} As for \pc{pathMatrix}, it is possible to specify a perfect orientation using \var{referencematching}.
\begin{flushleft}
\textit{For practical purposes, it is always computationally faster (and more general) to use \pc{pathMatrix}} to obtain a Grassmannian matrix associated to an on-shell diagram.
\end{flushleft}

Despite its complexity in comparison with \pc{pathMatrix}, this function has nonetheless been included in the package because signs are often of interest: for example, planar diagrams were discovered to belong to the \textit{positive} Grassmannian; also, signs play a determining role in \textit{regions} in the amplituhedron \cite{Arkani-Hamed:2013jha,Arkani-Hamed:2013kca,Franco:2014csa,Galloni:2016iuj} and may play an important role in constructing its triangulations.\footnote{Very interesting recent progress in the amplituhedron can additionally be found in \cite{Enciso:2014cta,Bai:2014cna,Lam:2014jda,Arkani-Hamed:2014dca,Bai:2015qoa,Ferro:2015grk,Bern:2015ple,Dennen:2016mdk,Arkani-Hamed:2016rak,Ferro:2016zmx,Ferro:2016ptt,Enciso:2016cif,Eden:2017fow}.}

The process of introducing signs in the boundary measurement requires the diagram to be embedded on a surface, with boundaries and cuts, cf.\ \cite{Franco:2013nwa}.\footnote{We note that this is true even for planar diagrams, which must be embedded on the disk before external nodes can be labeled cyclically.} \pc{grassmannianMatrix} makes all such choices automatically; moreover, for diagrams which can be embedded on a disk with no edge-crossings, it will always choose the standard cyclic ordering of external nodes, to ensure manifest positivity of the minors. It is worth noting that the automatically chosen ordering of external nodes will in general be different to the default ordering of \pc{pathMatrix}.
\begin{itemize}
\item[] 
{\small \textit{Example.} We shall now illustrate some of the differences between \pc{pathMatrix} and \pc{grassmannianMatrix} in a simple example, the square box, where in both cases we use the same perfect orientation by setting \var{referencematching} to \pc{Z[1]Z[2]}.
\begin{flalign*}
\incell{2} & \pc{pathMatrix[\{\{Z[1],Z[3]\},\{Z[4],Z[2]\}\},\{\{Z[5],0\},\{0,Z[6]\}\},} \\
& \quad \, \pc{\{\{Z[7],0\},\{0,Z[8]\}\},\{\{0,0\},\{0,0\}\},Z[1]Z[2]]}\:\pc{//}\:\pc{MatrixForm} \\
 & \\
& \pc{grassmannianMatrix[\{\{Z[1],Z[3]\},\{Z[4],Z[2]\}\},\{\{Z[5],0\},\{0,Z[6]\}\},} \\
& \quad \, \pc{\{\{Z[7],0\},\{0,Z[8]\}\},\{\{0,0\},\{0,0\}\},Z[1]Z[2]]}\:\pc{//}\:\pc{MatrixForm} \\
\outcell{2} & \left(
\begin{array}{cccc}
\pc{1}&\pc{0}&\frac{\pc{Z[5]}\pc{Z[7]}}{\pc{Z[1]}\left(1-\frac{\pc{Z[3]}\pc{Z[4]}}{\pc{Z[1]}\pc{Z[2]}}\right)}&\frac{\pc{Z[3]}\pc{Z[6]}\pc{Z[7]}}{\pc{Z[1]}\pc{Z[2]}\left(1-\frac{\pc{Z[3]}\pc{Z[4]}}{\pc{Z[1]}\pc{Z[2]}}\right)} \\
\pc{0}&\pc{1}&\frac{\pc{Z[4]}\pc{Z[5]}\pc{Z[8]}}{\pc{Z[1]}\pc{Z[2]}\left(1-\frac{\pc{Z[3]}\pc{Z[4]}}{\pc{Z[1]}\pc{Z[2]}}\right)}&\frac{\pc{Z[6]}\pc{Z[8]}}{\pc{Z[2]}\left(1-\frac{\pc{Z[3]}\pc{Z[4]}}{\pc{Z[1]}\pc{Z[2]}}\right)} \\
\end{array}
\right) \\
\outcell{3} & \left(
\begin{array}{cccc}
\pc{1}&\frac{\pc{Z[5]}\pc{Z[7]}}{\pc{Z[1]}\left(1+\frac{\pc{Z[3]}\pc{Z[4]}}{\pc{Z[1]}\pc{Z[2]}}\right)}&\pc{0}&-\frac{\pc{Z[3]}\pc{Z[6]}\pc{Z[7]}}{\pc{Z[1]}\pc{Z[2]}\left(1+\frac{\pc{Z[3]}\pc{Z[4]}}{\pc{Z[1]}\pc{Z[2]}}\right)} \\
\pc{0}&\frac{\pc{Z[4]}\pc{Z[5]}\pc{Z[8]}}{\pc{Z[1]}\pc{Z[2]}\left(1+\frac{\pc{Z[3]}\pc{Z[4]}}{\pc{Z[1]}\pc{Z[2]}}\right)}&\pc{1}&\frac{\pc{Z[6]}\pc{Z[8]}}{\pc{Z[2]}\left(1+\frac{\pc{Z[3]}\pc{Z[4]}}{\pc{Z[1]}\pc{Z[2]}}\right)} \\
\end{array}
\right)
\end{flalign*}
Here we can explicitly see two features of \pc{grassmannianMatrix}: the ordering of external nodes is different to \pc{pathMatrix}, and its signs have been changed in order to ensure manifest positivity of the minors, for positive edge weights. The precise ordering of nodes in \pc{grassmannianMatrix} can be found out using the function \newline \pc{getOrderingExternalNodesGrassmannian}, introduced in \sref{sec:scatteringinfofunctions}; using \pc{turnIntoGraph} it is then possible to explicitly verify that \pc{grassmannianMatrix} has indeed chosen a cyclic ordering of nodes on the disk.
}%
\end{itemize}

While fully automatic, this function has been designed to permit the user to take control of the choice of embedding, with boundaries and cuts, and input a user-specified ordering of external nodes. We remind the reader that the chosen ordering of nodes must be consistent with the chosen embedding, as we shall discuss in the next example.
\begin{itemize}
\item[] 
{\small \textit{Example.} We shall illustrate the format for how to specify the embedding using the example shown in \fref{fig:externalnodeordering}. In the figure all external nodes lie on one of the $B=3$ different boundaries; the $B-1$ cuts between the boundaries are shown as dotted green lines. As explained in detail in \cite{Franco:2013nwa}, the ordering of external nodes must follow the chosen boundaries and cuts, similarly to complex analysis, as shown by the yellow line and arrows; this chronological ordering along the yellow path is shown using the green number labels. The blue numbers show the node numbers of the graph, according to the Kasteleyn matrix (which counts white nodes first followed by black nodes). We stress that these are the node numbers of the diagram, and have nothing to do with the embedding-dependent \textit{ordering} of the external nodes. The edges crossed by the cuts have been labeled in red.
\begin{figure}[hbt]
\begin{center}
\includegraphics[scale=0.3]{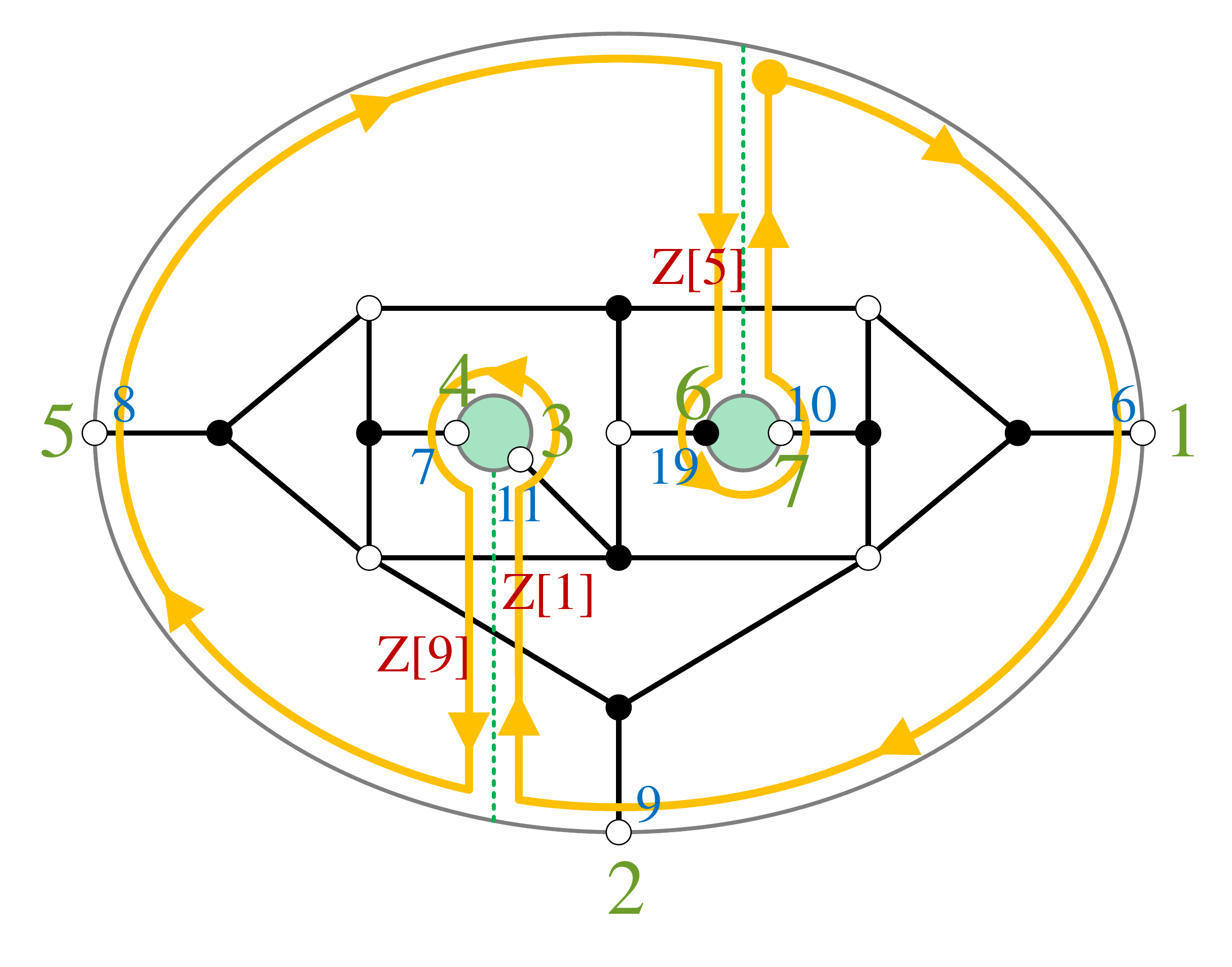}
\vspace{-0.75cm}
\caption{}
\label{fig:externalnodeordering}
\end{center}
\end{figure}

In \pc{grassmannianMatrix} the parameters which specify the embedding are \var{externalordering}, \var{boundarylist} and \var{boundarycutreplacements}:
\begin{itemize}
\item[$\bullet$] \var{externalordering} is given as a list of node numbers, appearing in the correct order given by the cuts and boundaries; for this example it would be \pc{\{6,9,11,7,8,19,10\}}.
\item[$\bullet$] \var{boundarylist} is a list of boundaries, where each boundary is given as a list of nodes on that boundary. For this example \var{boundarylist} is \pc{\{\{6,9,8\},\{11,7\},\{19,10\}\}}.
\item[$\bullet$] When a cut crosses an edge, the edge picks up a minus sign. For this reason, the cuts are specified by which pair of boundaries they connect, and which edges they cross. For this example \var{boundarycutreplacements} is \newline \pc{\{\{\{6,9,8\},\{11,7\}\}$\to$\{Z[1]$\to -$Z[1],Z[9]$\to -$Z[9]\},}\newline \pc{ \{\{6,9,8\},\{19,10\}\}$\to$\{Z[5]$\to -$Z[5]\}\}} \newline
which indicates that the cut connecting the boundaries \pc{\{6,9,8\}} and \pc{\{11,7\}} forces \pc{Z[1]} and \pc{Z[9]} to pick up a sign, and the cut connecting the boundaries \pc{\{6,9,8\}} and \pc{\{19,10\}} forces \pc{Z[5]} to pick up a sign.
\end{itemize}
}%
\end{itemize}

\pointscat \func{dimensionGrassmannian}{\kast}: Returns the dimension of the Grassmannian associated to the on-shell diagram given by the Kasteleyn matrix. This is computed as the dimension of the tangent space of its \pl coordinates, which in turn are expressed in the gauge-unfixed form given by \pc{minorsAsPerfectMatchings} described below. We stress that \pc{dimensionGrassmannian} works very well also in the presence of relations among \pl coordinates which are independent from the \pl relations, as often happens for non-planar diagrams.

\pointscat \func{pluckerCoordinates}{\kast,\newline \phantom{444444}\varbardef{referencematching}{Null},\varbardef{withsigns}{False}}: Returns the minors of the Grassmannian matrix associated to the on-shell diagram. As for \pc{pathMatrix} and \pc{grassmannianMatrix}, it is possible to choose a particular perfect orientation by setting the optional input \var{referencematching} as the perfect matching corresponding to the chosen perfect orientation. By default this function will use the path matrix given by \pc{pathMatrix}; to use the matrix given by \pc{grassmannianMatrix} it is necessary to set the optional input \var{withsigns} to \pc{True}.

\pointscat \func{minorsAsPerfectMatchings}{\kast,\newline \phantom{444444}\varbardef{referencematching}{Null}}: A necessary requirement for a valid map from an on-shell diagram to an element of the Grassmannian is that the \pl coordinates simplify to a sum of perfect matchings, where each perfect matching in the sum has its corresponding source set equal to the label of the \pl coordinate \cite{Franco:2013nwa,Franco:2015rma}. Choosing a perfect orientation gauge-fixes the Grassmannian matrix such that some \pl coordinate is equal to 1. The function \pc{minorsAsPerfectMatchings} restores the gauge freedom, and expresses all \pl coordinates as their true sum of perfect matchings (without including any manifest-positivity signs in \pc{grassmannianMatrix}).
\begin{flushleft}
The function \pc{minorsAsPerfectMatchings} is (currently) the best way to express the d.o.f.\ of the on-shell diagram in a form which is manifestly Grassmannian, and in particular \textit{is superior to using gauge-fixed \pl coordinates}.
\end{flushleft}
\pc{minorsAsPerfectMatchings} has another very strong advantage---boundaries of the Grassmannian can be easily obtained \textit{without ever having to worry about coordinate charts}: simply setting \textit{any} removable edge to zero will neatly kill those perfect matchings which utilize that edge, immediately yielding the new gauge-free \pl coordinates.

The ordering of the \pl coordinates in the list given by \pc{minorsAsPerfectMatchings} is lexicographic, i.e.\ $\{\Delta_{12 \cdots k}$, $\Delta_{12 \cdots (k-1) (k+1)}$, $ \ldots, \Delta_{(n-k) \cdots n}\}$. The ordering of external nodes is the same as that of \pc{pathMatrix}, i.e.\ it begins with the rows of \var{bottomleft} and continues with by the columns of \var{topright}. 

\subsection[Properties of \N{4} On-Shell Diagrams]{Properties of $\boldsymbol{\mathcal{N}=4}$ On-Shell Diagrams}
\label{sec:scatteringinfofunctions}

The functions in \sref{sec:grassmannianfunctions} were primarily aimed at obtaining Grassmannian-like quantities related to each on-shell diagram. This section will present functions that establish various properties of on-shell diagrams.

\pointscat \func{planarityQ}{\kast}: Returns \pc{True} if the diagram is ``planar'', i.e.\ can be embedded on a disk with no edges crossing,\footnote{It is possible to have a diagram which is ``planar'' and yet not be embeddable on the disk; this is only known to happen for diagrams whose bivalent nodes have not been collapsed. For this reason, \pc{planarityQ} automatically collapses bivalent nodes before checking whether there exists a disk embedding. For the purposes of scattering amplitudes, diagrams that are move-equivalent to planar diagrams should also count as planar. The function \pc{grassmannianMatrix} cannot account for this, however, and will order external nodes according to whichever embedding is possible on the given diagram, \textit{without} manipulations such as collapsing bivalent nodes.} and \pc{False} otherwise. We highlight the difference between this function and the in-built Mathematica function \pc{PlanarGraphQ}, which tells whether the diagram can  or cannot be embedded on genus zero without edges crossing.

\pointscat \func{reducibilityQ}{\kast}: Returns \pc{True} if the diagram is reducible, i.e.\ if there are edges which can be removed without affecting the dimensionality of the Grassmannian associated to the diagram, and \pc{False} if the diagram is reduced.

\pointscat \func{reducibilityEdges}{\kast}:  Returns a list of edges which, if removed, do not change the dimension of the Grassmannian associated to the diagram. It is not possible to simultaneously remove all edges in the returned list; rather, the removal of any \textit{one} edge from this list will not decrease the dimension of the Grassmannian. To achieve a full reduction it is necessary to use the function \pc{reductionGraph}, described below.

\pointscat \func{reductionGraph}{\kast}: Returns all possible sets of edges which, if \textit{all} edges in a given set are removed, do not decrease the dimension of the Grassmannian. Not all sets will achieve a maximal reduction; the list given by \pc{reductionGraph} is ordered with smaller reductions first, and ends with all maximal reductions.

\pointscat \func{getK}{\kast}: Returns the number $k$ describing the helicities of external particles in a N$^{k-2}$MHV on-shell diagram. This number also equals the number of sources in any perfect orientation of the diagram.

\pointscat \func{getN}{\kast}: Returns the number of external particles in an on-shell diagram, which equals the number of external nodes in the Kasteleyn.

\pointscat \func{getSourceNodes}{\kast,\varbar{referenceperfmatch}}: Each perfect orientation associated to a perfect matching has a set of sources and sinks. Given a Kasteleyn and a perfect matching \var{referenceperfmatch}, this function will return a list of node numbers which are sources in the corresponding perfect orientation. We remind the reader that the numbering of nodes of the diagram follows the order of rows and columns in the Kasteleyn, starting with internal white nodes, followed by external white nodes, internal black nodes, and finally external black nodes.

\pointscat \func{getSinkNodes}{\kast,\varbar{referenceperfmatch}}: Returns a list of node numbers which are sinks in the corresponding perfect orientation. The numbering of nodes of the diagram is the same as for \pc{getSourceNodes}.

\pointscat \func{getOrderingExternalNodesDefault}{\kast}: Returns an ordered list of node numbers, ordered according to the automatic default ordering of external nodes, used by e.g.\ \pc{pathMatrix}. This ordering begins with the node numbers representing the rows of \var{bottomleft} and continues with the node numbers representing the columns of \var{topright}.
\begin{itemize}
\item[] 
{\small \textit{Example.} For the example of the square box used to illustrate the function \pc{grassmannianMatrix}, \pc{getOrderingExternalNodesDefault} returns \pc{\{3,4,7,8\}}, indicating that node number 3 in the diagram is the first external node and node number 8 is the last external node, in the default ordering.
}%
\end{itemize}

\pointscat \func{getOrderingExternalNodesGrassmannian}{\kast}: Returns an ordered list of node numbers, ordered according to the algorithmically chosen ordering of external nodes used by \pc{grassmannianMatrix}, which will in general be different to that returned by \pc{getOrderingExternalNodesDefault}. For diagrams embeddable on the disk, this will be a cyclic ordering of nodes around the disk; for non-planar diagrams the ordering will be chosen using an algorithm which facilitates an automatic choice of cuts.
\begin{itemize}
\item[] 
{\small \textit{Example.} For the example of the square box used to illustrate the function \pc{grassmannianMatrix}, \pc{getOrderingExternalNodesGrassmannian} returns \pc{\{3,7,4,8\}}, i.e.\ it has swapped the second and third nodes with respect to the default ordering. Indeed, using \pc{turnIntoGraph} it is easy to see that this new ordering is a cyclic ordering when the diagram is embedded on the disk.
}%
\end{itemize}

\pointscat \func{getExternalEdgeNodeNumbers}{\kast,\newline \phantom{444444}\varbar{externaledgelist}}: It is often useful to translate the names of external edges into external node numbers. This function returns the external node numbers associated to the external edges in the list \var{externaledgelist}. The node numbers are always returned in numeric order, i.e.\ they do \textit{not} preserve the order of edges in \var{externaledgelist}.

\pointscat \func{getSourceEdges}{\varbar{topright},\varbar{bottomleft},\varbar{referenceperfmatch}}: For complete, connected on-shell diagrams,\footnote{Which are most diagrams considered in scattering amplitudes which are not boundaries in a stratification.} all external nodes are the end point of an edge, i.e.\ no external nodes are disconnected from the graph. Thus, it can be useful to talk about external edges rather than external nodes; this function returns the list of ``source edges'' for the perfect orientation corresponding to the perfect matching \var{referenceperfmatch}. We note that this function does not require as input the entire Kasteleyn matrix.

\pointscat \func{getSinkEdges}{\varbar{topright},\varbar{bottomleft},\varbar{referenceperfmatch}}: Similarly to \pc{getSourceEdges}, this function will return the ``sink edges'' for the perfect orientation corresponding to the perfect matching \var{referenceperfmatch}. We note that this function does not require as input the entire Kasteleyn matrix.

\pointscat \func{getOrderingExternalEdgesDefault}{\kast}: Returns a list of replacement rules, where each entry is a \pc{Rule} between an external edge and its ordering number, as ordered by \pc{pathMatrix}. We stress that the edges are \textit{not} replaced by their node numbers, but rather by the ordering of those node numbers  in \pc{getOrderingExternalNodesDefault}.
\begin{itemize}
\item[] 
{\small \textit{Example.} For the example of the square box used to illustrate the function \pc{grassmannianMatrix}, \pc{getOrderingExternalEdgesDefault} returns \pc{\{Z[7]}$\to$\pc{1,Z[8]}$\to$\pc{2,Z[5]}$\to$\pc{3,Z[6]}$\to$\pc{4}\pc{\}}, indicating that the default ordering of external edges places \pc{Z[7]} as the first external edge and \pc{Z[6]} as the last external edge.
}%
\end{itemize}

\pointscat \func{getOrderingExternalEdgesGrassmannian}{\kast}: Returns a list of replacement rules, where each entry is a \pc{Rule} between an external edge and its ordering number, as ordered by \pc{grassmannianMatrix}. We stress that the edges are \textit{not} replaced by their node numbers, but rather by the ordering of those node numbers  in \pc{getOrderingExternalNodesGrassmannian}.
\begin{itemize}
\item[] 
{\small \textit{Example.} For the example of the square box used to illustrate the function \pc{grassmannianMatrix}, \pc{getOrderingExternalEdgesGrassmannian} returns \pc{\{Z[7]}$\to$\pc{1,Z[5]}$\to$\pc{2,Z[8]}$\to$\pc{3,Z[6]}$\to$\pc{4}\pc{\}}, indicating that \pc{Z[5]} is the second external edge in the ordering chosen by \pc{grassmannianMatrix} and \pc{Z[8]} is the third external edge in this ordering, contrary to \pc{getOrderingExternalEdgesDefault}.
}%
\end{itemize}

\subsection{The Stratification of On-Shell Diagrams}
\label{sec:stratificationfunctions}

Many recent developments in $\mathcal{N}=4$ scattering amplitudes have focused on the study of the \textit{stratification} of on-shell varieties.\footnote{An on-shell variety is the space spanned by the degrees of freedom of an on-shell diagram.} The geometric boundary structure of these regions in the Grassmannian perfectly mimics the singularity structure of the amplitude integrand associated to the on-shell diagram. Recent tools \cite{Bourjaily:2016mnp} have enabled a completely general approach to this stratification. The functions presented in this section are aimed at simplifying access to these tools to aid future exploration of on-shell diagrams.

\pointscat \func{removableEdges}{\kast}: Returns a list of equivalence classes of edges;\footnote{Removing a single edge will occasionally cause other edges in the diagram to not participate in any perfect matching, and hence be irrelevant to the diagram. These edges must be removed simultaneously with the original edge \cite{2007arXiv0706.2501P,Franco:2013nwa}. Together, this set of edges is known as an equivalence class of edges.} removing all the edges in any one such equivalence class will access a boundary. For planar diagrams these boundaries are all distinct; in non-planar diagrams this is not true in general and it may be possible to access the same boundary using either one of different equivalence classes of edges \cite{Bourjaily:2016mnp}.

\pointscat \func{stratificationBoundaries}{\kast}: Computes and returns the full stratification of arbitrary on-shell diagrams, which is conjectured to be equal to the singularity structure of the associated on-shell form. The method for computing the stratification is an efficient version of the tools presented in \cite{Bourjaily:2016mnp}. The format for how \pc{stratificationBoundaries} returns the stratification structure is as follows: it is a list, where each element of the list is a list of codimension-$i$ boundaries. The first element given by \pc{stratificationBoundaries} contains the list of codimension-0 boundaries (there is only one element here, the top-dimensional one) and the last element contains the zero-dimensional boundaries.

When computing the stratification, we remove all possible removable edges from each boundary and sub-boundary. Some of the resulting configurations will be equivalent, i.e.\ it is often possible to remove different sets of edges from the top-dimensional on-shell diagram and end up with the same on-shell variety. For this reason, each boundary given by \pc{stratificationBoundaries} is represented as a list of equivalent configurations of on-shell diagrams, where each configuration is a list of those edges which are present in the diagram.
\begin{itemize}
\item[] 
{\small \textit{Example.} Since the nested lists of \pc{stratificationBoundaries} may at first appear hard to visualize, let us consider an explicit, very simple example of such a stratification, namely the non-planar example also used to illustrate the functions \pc{pathMatrix} and \pc{loopDenominator}.
\begin{flalign*}
\incell{2} & \pc{stratification=stratificationBoundaries[\{\{Z[1],Z[2],Z[3],0\},} \\
& \quad \; \; \pc{\{Z[4],Z[5],0,Z[6]\},\{0,0,Z[7],Z[8]\}\},\{\{0\},\{0\},\{Z[9]\}\},} \\
& \quad \; \; \pc{\{\{Z[10],0,0,0\},\{0,Z[11],0,0\},\{0,0,Z[12],0\},\{0,0,0,Z[13]\}\},} \\
& \quad \; \; \pc{\{\{0\},\{0\},\{0\},\{0\}\}];} \\
 & \pc{stratification[[2]]} \\
 & \pc{stratification[[6]][[1]]} \\
\outcell{3} & \pc{\{\{\{Z[2],Z[3],Z[4],$\cdots$,Z[11],Z[12],Z[13]\}\},} \\
& \; \; \pc{\{\{Z[1],Z[3],Z[4],$\cdots$,Z[11],Z[12],Z[13]\}\},$\cdots$} \\
& \; \; \pc{$\cdots$,\{\{Z[1],Z[2],Z[3],$\cdots$,Z[11],Z[12],Z[13]\}\}\}} \\
\outcell{4} & \pc{\{\{Z[2],Z[4],Z[8],Z[9],Z[12],Z[13]\},} \\
& \; \; \pc{\{Z[1],Z[5],Z[8],Z[9],Z[12],Z[13]\}\}}
\end{flalign*}
The output of \pc{stratification[[2]]} is longer than shown above and was shortened with dots to illustrate the structure of the output more transparently. \pc{stratification[[2]]} is the second element of the stratification, i.e.\ the collection of all codimension-1 boundaries. \pc{Length[stratification[[2]]]} equals 6, telling us that there are 6 distinct codimension-1 boundaries. Moreover, each boundary is a list of equivalent configurations that represent this boundary. In the case at hand, all boundaries in \pc{stratification[[2]]} are represented by a single configuration; this is a reflection of the fact that from the top-dimensional on-shell diagram, different removable edges access different boundaries.\footnote{As already mentioned, it is possible in general to have different removable edges access the same codimension-1 boundary. This would manifest itself above as some element in \pc{stratification[[2]]} whose length is greater than 1.} Each configuration contains all edges except for one, which is the removable edge used to access this boundary.

We also look at the first of the codimension-5 boundaries, i.e.\ \pc{stratification[[6]][[1]]}. Here we see that there are two different but equivalent configurations representing the boundary: the diagram with edges \pc{\{Z[2],Z[4],Z[8],Z[9],Z[12],Z[13]\}} and the diagram with edges \pc{\{Z[1],Z[5],Z[8],Z[9],Z[12],Z[13]\}}, where all other edges have been removed.
}%
\end{itemize}
Finally, we remark on what happens when applying \pc{stratificationBoundaries} to reducible diagrams. In these cases, the function will remove all possible combinations of edges until it obtains reduced diagrams. It will then proceed with only removing removable edges from these diagrams. In the output of \pc{stratificationBoundaries} the list of codimension-0 objects will be  an empty list. This reflects the fact that without removing any edges, we have no reduced on-shell diagrams. If after removing any one edge, there are still no reduced on-shell diagrams, the second element of the output of \pc{stratificationBoundaries} will also be an empty list. Eventually, after having removed enough edges, we will obtain a reduced on-shell diagram; this will be the first non-empty list returned by \pc{stratificationBoundaries}.\footnote{The number of empty lists does not necessarily equal the number of removed edges; it is equal to the number of loop-variable degrees of freedom lost by the removal of some set of edges. Expressed differently, it equals the difference in dimensionality of the matching polytope, between the original diagram and the sub-diagram obtained by removing this set of edges.} It may be that there are multiple distinct boundaries in the first non-empty list; this is due to the fact that there often exist several inequivalent maximal reductions of a reducible diagram.

\pointscat \func{stratificationNumbers}{\kast}: Returns a list containing the total \textit{number} of codimension-$i$ boundaries, where the list is ordered starting from highest dimension and ending with zero-dimensional boundaries. If the starting diagram is reducible, the first element of the list will be a zero, following the discussion for the function \pc{stratificationBoundaries}. 
\begin{itemize}
\item[] 
{\small \textit{Example.} We shall apply \pc{stratificationNumbers} to the same example as that used to illustrate the function \pc{stratificationBoundaries}.
\begin{flalign*}
\incell{2} & \pc{stratificationNumbers[\{\{Z[1],Z[2],Z[3],0\},\{Z[4],Z[5],0,Z[6]\},} \\
& \quad \; \; \pc{\{0,0,Z[7],Z[8]\}\},\{\{0\},\{0\},\{Z[9]\}\},\{\{Z[10],0,0,0\},} \\
& \quad \; \; \pc{\{0,Z[11],0,0\},\{0,0,Z[12],0\},\{0,0,0,Z[13]\}\},} \\
& \quad \; \; \pc{\{\{0\},\{0\},\{0\},\{0\}\}]} \\
\outcell{2} & \pc{\{1,6,21,35,42,30,10\}}
\end{flalign*}
From this we see that there is a single top-dimensional element, namely the on-shell diagram we gave to the function, and 6 codimension-1 boundaries, as already noted in the example for \pc{stratificationBoundaries}. Finally,  we have 10 zero-dimensional boundaries, which are simply the 10 different \pl coordinates of G(3,5).
}%
\end{itemize}
We remark that it is possible, however rare, for diagrams to have the same stratification numbers without being related by equivalence moves, a reordering of external nodes, or color swaps \cite{Bourjaily:2016mnp}. Hence, the list given by \pc{stratificationNumbers} is not a full diagnostic for categorizing on-shell varieties.

\pointscat \func{stratificationEulerNumber}{\kast}: Returns the Euler number of the stratification, by first computing the numbers of boundaries of each dimension using \pc{stratificationNumbers} and then computing the alternating sum of these numbers of boundaries.
\begin{itemize}
\item[] 
{\small \textit{Example.} Applying \pc{stratificationEulerNumber} to the same example as that used to illustrate the functions \pc{stratificationNumbers} and \pc{stratificationBoundaries}, we obtain
\begin{flalign*}
\incell{2} & \pc{stratificationEulerNumber[\{\{Z[1],Z[2],Z[3],0\},\{Z[4],Z[5],0,Z[6]\},} \\
& \quad \; \; \pc{\{0,0,Z[7],Z[8]\}\},\{\{0\},\{0\},\{Z[9]\}\},\{\{Z[10],0,0,0\},} \\
& \quad \; \; \pc{\{0,Z[11],0,0\},\{0,0,Z[12],0\},\{0,0,0,Z[13]\}\},} \\
& \quad \; \; \pc{\{\{0\},\{0\},\{0\},\{0\}\}]} \\
\outcell{2} & \pc{3}
\end{flalign*}
All planar diagrams have Euler number equal to 1; this result is a clear indicator of the non-planarity of the graph in our example.
}%
\end{itemize}

\pointscat \func{stratificationGraph}{\kast}: Returns the graph of the stratification, i.e.\ a graph where nodes represent distinct boundaries and outgoing arrows from nodes represent the boundaries of these nodes. As for \pc{matchingPolytopeBoundariesGraph}, Mathematica may not always choose to embed the stratification graph in a layered way, where each layer corresponds to a different dimensionality; to obtain a graph embedded in this form, it is necessary to evaluate the graph with the option \pc{Graph[graph, GraphLayout$\mathtt{\to}$"{\color[RGB]{102,102,102} LayeredDigraphEmbedding}"]}, where \pc{graph} is the variable that stores the graph given by \pc{stratificationGraph}. The function works well also for reducible on-shell diagrams.
\begin{itemize}
\item[] {\small \textit{Example.} We shall illustrate \pc{stratificationGraph} for the simple example of the square box, whose graph of boundaries easily fits on a page. 
\begin{figure}[hbt]
\begin{center}
\includegraphics[width=\textwidth]{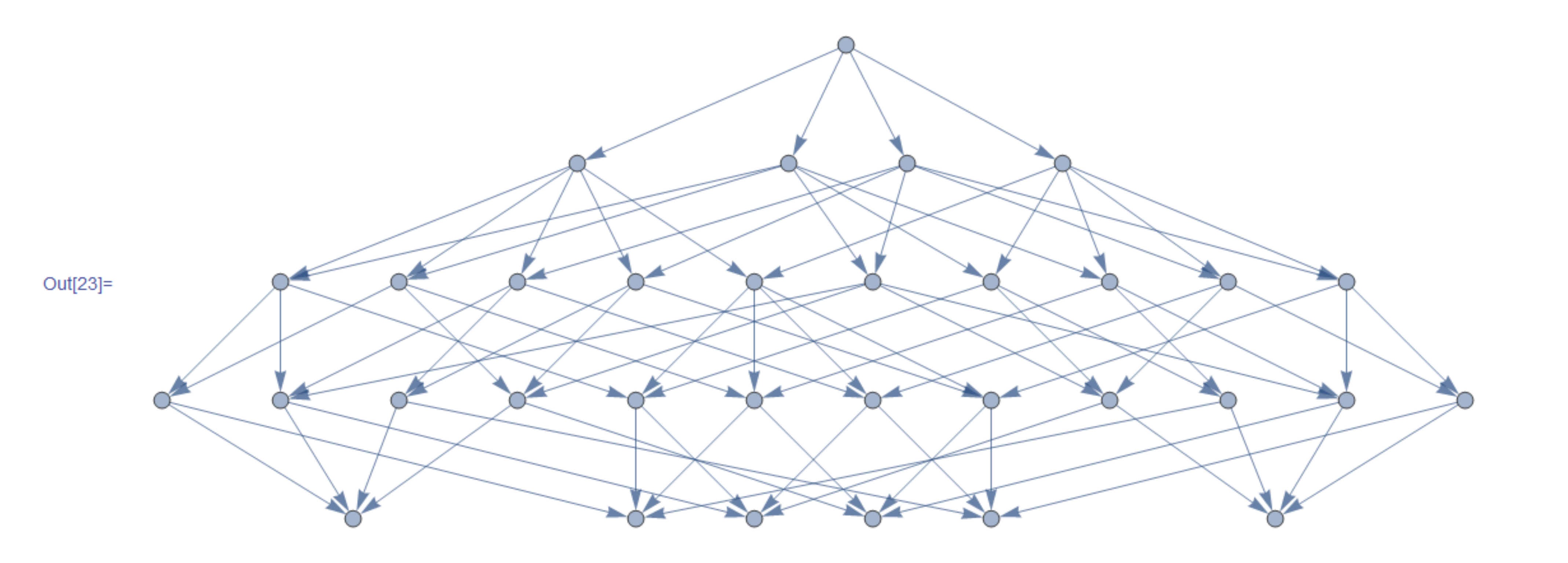}
\label{fig:stratGraphSqb}
\end{center}
\end{figure}
}%
\end{itemize}
Two reduced diagrams are related by equivalence moves (such as square moves, bubble reduction, and collapsing bivalent vertices), reordering of external nodes, and color swap (i.e.\ parity conjugation) if and only if their stratification graphs are isomorphic. This can easily be checked with Mathematica using the in-built function \pc{IsomorphicGraphQ}. This is the safest way to fully categorize distinct on-shell varieties.

\pointscat \func{nonPluckerPolesQ}{\kast}: Determines whether the on-shell form corresponding to the on-shell diagram has poles which are not expressed as \pl coordinates, but rather are expressed as more complicated expressions of \pl coordinates. The function returns \pc{True} if such complicated poles exist and \pc{False} otherwise. These ``non-standard poles'' are a new phenomenon of non-planar diagrams; however, not all non-planar diagrams display this phenomenon. 

\subsection{\pl Coordinates and Matroids}
\label{sec:mathgrassmannianfunctions}

\vspace{-0.3cm}
\pointscat \func{matroidQ}{\varbar{inputmatroid}}: A matroid is a set of $k$-element subsets of a collection of $n$ elements which satisfies the ``exchange axiom''. The function \pc{matroidQ} accepts as input a list of $k$-element lists. It then returns \pc{True} if the input satisfies the exchange axiom, and \pc{False} otherwise.
\begin{itemize}
\item[] 
{\small \textit{Example.} Here we give an example of a set of 3-element lists that is a matroid and one that is not.
\begin{flalign*}
\incell{2} & \pc{matroidQ[\{\{1,2,3\},\{1,2,4\},\{1,2,5\},\{1,2,6\}\}]} \\
& \pc{matroidQ[\{\{1,2,3\},\{1,2,4\},\{1,2,5\},\{1,3,4\}\}]} \\
\outcell{2} & \pc{True} \\
\outcell{3} & \pc{False}
\end{flalign*}
}%
\end{itemize}

\pointscat \func{matroidViolationCheck}{\varbar{inputmatroid}}: Takes as input a list of $k$-element lists, similarly to the function \pc{matroidQ}, and returns a list of pairs of matroids that do not satisfy the exchange axiom. If \var{inputmatroid} is a matroid it will return an empty list.
\begin{itemize}
\item[] 
{\small \textit{Example.} Here we consider the same example as for \pc{matroidQ}.
\begin{flalign*}
\incell{2} & \pc{matroidViolationCheck[\{\{1,2,3\},\{1,2,4\},\{1,2,5\},\{1,2,6\}\}]} \\
& \pc{matroidViolationCheck[\{\{1,2,3\},\{1,2,4\},\{1,2,5\},\{1,3,4\}\}]} \\
\outcell{2} & \pc{\{\}} \\
\outcell{3} & \pc{\{\{\{1,2,5\},\{1,3,4\}\}\}}
\end{flalign*}
Here we both see that the first case is a matroid and the second one is not, but also that in the second case there is a single pair of elements that does not satisfy the exchange axiom, namely \pc{\{1,2,5\}} and \pc{\{1,3,4\}}: removing \pc{2} from the first set and replacing it with any of the elements in \pc{\{1,3,4\}} does not create a set which appears in the matroid, thus violating the exchange axiom.
}%
\end{itemize}

\pointscat \func{pluckerRelations}{\varbar{k},\varbar{n}}: Returns a list of \pl relations for a given $n$ number of external particles in a N$^{k-2}$MHV process. If the Mathematica file \pc{premadePluckerRelations} is placed in the working directory of the Mathematica session (found using the in-built function \pc{Directory[]}), \pc{pluckerRelations} will import the answer for cases with larger $k$ and $n$, which would otherwise take a long time to compute. Not all the \pl relations given by \pc{pluckerRelations} are independent however; the use of this function is intended as an aid when using \pl relations to simplify complicated expressions. 

Each \pl coordinate has the form \pc{HoldForm[minor][$indices$]}, which in a Mathematica notebook displays simply as \pc{minor[$indices$]}. The \pc{HoldForm} was put in place in case the user had already specified a value for some variable called \pc{minor}; to remove the \pc{HoldForm} it is necessary to use the in-built function \pc{ReleaseHold} on the expressions given by \pc{pluckerRelations}.

\pointscat \func{independentPluckerRelations}{\varbar{k},\varbar{n}}: It is often very useful to have a minimal set of \pl relations, and solve them, in order to check the veracity of various relations among \pl coordinates. In particular, it is sometimes necessary to verify whether some particular relation follows from the \pl relations or whether it is independent of them. \pc{independentPluckerRelations} returns a list containing two elements: the first is a list of independent \pl relations; the second is a solution to the \pl relations. As for \pc{pluckerRelations}, this function will be faster for larger $k$ and $n$ if the file \pc{premadePluckerRelations} is placed in the working directory of the Mathematica session. Also, \pc{HoldForm} is wrapped around the symbol \pc{minor} to protect it from any previous definitions of the variable name.

\subsection[\N{1} Bipartite Field Theories]{$\boldsymbol{\mathcal{N}=1}$ Bipartite Field Theories}
\label{sec:bftfunctions}

Bipartite Field Theories have a host of tools available for computing quantities of interest, such as their moduli space or their zig-zag paths. The functions in this section summarize some of the most common tools used, whose computation utilizes the most efficient approach which is currently known. As mentioned in \sref{sec:bftembedding}, gauging 1 depends on the embedding, which is specified through the labeling of indices of the edges, while gauging 2 does not require this information. However, in the context of BFTs it is often useful to discuss the theory in terms of an embedding regardless of which gauging one is interested in, and it is recommended to label edges accordingly. The functions in this section do not require such a labeling for gauging 2, but certain functions cannot possibly give a sensible answer without an embedding, e.g.\ those involving zig-zag paths, which are highly embedding-dependent.

\pointbft \func{moduliSpaceBFT}{\kast,\varbar{gauging}}: Returns a matrix representing the toric diagram of the moduli space of the BFT given by the Kasteleyn matrix. Each column of the matrix represents the coordinate in the toric diagram of each perfect matching. The columns appear in the same order as the corresponding perfect matchings in the function \pc{perfectMatchings}. Setting the parameter \var{gauging} to \pc{2} will return the matroid polytope as given by \pc{matroidPolytope} \cite{Franco:2012wv}. The coordinates may in reality describe a polytope of lower dimension than the number of rows in the matrix; to view the actual dimensionality of the polytope, simply use the function \pc{dimensionPolytope} introduced in \sref{sec:generalbipartitafunctions}.

\pointbft \func{moduliLoopVariablesBFT}{\kast,\varbar{gauging},\newline \phantom{444444}\varbardef{referencematching}{Null},\varbardef{loopvariablebasis}{Null}}: Computes the moduli space using generalized loop variables \cite{Franco:2012wv}. In a given perfect orientation, each perfect matching corresponds to a specific flow, whose precise formulation is given by dividing the perfect matching by the reference perfect matching\footnote{The reference perfect matching is the one that gives rise to the chosen perfect orientation.}. Edges appearing in the numerator are directed from white node to black node and edges appearing in the denominator are oppositely oriented. 

Each flow can be written as a product of loop variables; in this way, it is possible to relate each perfect matching to a specific product of loop variables, which in turn can be specified by the powers of each of the loop variables. The toric diagram coordinate for each perfect matching is given by the set of powers of the loop variables used to describe its flow in the given perfect orientation \cite{Franco:2012wv}.

The function \pc{moduliLoopVariablesBFT} returns a list of three elements: the first is the toric diagram coordinates of the master space, in the form of a matrix whose columns correspond to perfect matchings and rows correspond to powers of loop variables. The second element in the list returned by \pc{moduliLoopVariablesBFT} is the toric diagram coordinates for the moduli space of the BFT, which is obtained by gauging away loop degrees of freedom associated to gauge groups; the mandatory input \var{gauging} should be set to \pc{1} or \pc{2} depending on what gauging the user is interested in. The third element of the list is the loop variable basis used in the rows of the master space given in the first element of the list.

There are two additional optional inputs: the user may specify the reference perfect matching, which determines the perfect orientation, with \var{referencematching}. It is also possible to specify the specific loop variable basis by setting \var{loopvariablebasis} to a list where each element is a loop variable (written as a product of edges, appearing in the numerator or denominator depending on their orientation).
\begin{itemize}
\item[] 
{\small \textit{Example.} We shall consider a simple example of a BFT embedded on genus zero with two boundaries. There is only one internal face, here denoted \pc{1}.
\begin{flalign*}
\incell{2} & \pc{topleft=\{\{X[6,2],X[2,1],X[1,6],0\},\{X[3,6],X[1,3],0,X[6,1]\},} \\
& \quad \; \; \pc{\{0,0,X[4,1],X[1,5]\}\};} \\
& \pc{topright=\{\{0\},\{0\},\{X[5,4]\}\};} \\
& \pc{bottomleft=\{\{X[2,3],0,0,0\},\{0,X[3,2],0,0\},\{0,0,X[6,4],0\},} \\
& \quad \; \; \pc{\{0,0,0,X[5,6]\}\};} \\
& \pc{bottomright=\{\{0\},\{0\},\{0\},\{0\}\};}
\end{flalign*}
We shall compute the master and moduli spaces using gauging 1.
\begin{flalign*}
\incell{6} & \pc{\{master,moduli,basis\}=moduliLoopVariablesBFT[topleft,topright,} \\
& \quad \; \; \pc{bottomleft,bottomright,1];} \\
& \pc{master}\:\pc{//}\:\pc{MatrixForm} \\
& \pc{moduli}\:\pc{//}\:\pc{MatrixForm} \\
& \pc{basis} \\
\outcell{7} & \left(
\begin{array}{ccccccccccccccc}
\pc{0} & \pc{0} & \pc{0} & \pc{0} & \pc{0} & \pc{1} & \pc{1} & \pc{0} & \pc{1} & \pc{-1} & \pc{-1} & \pc{0} & \pc{-1} & \pc{-1} & \pc{0} \\
\pc{0} & \pc{-1} & \pc{-1} & \pc{0} & \pc{-1} & \pc{0} & \pc{0} & \pc{0} & \pc{0} & \pc{-2} & \pc{-2} & \pc{-1} & \pc{-1} & \pc{-1} & \pc{-1} \\
\pc{0} & \pc{0} & \pc{-1} & \pc{-1} & \pc{-1} & \pc{0} & \pc{0} & \pc{-1} & \pc{0} & \pc{-1} & \pc{-2} & \pc{-1} & \pc{-1} & \pc{-2} & \pc{-1} \\
\pc{0} & \pc{0} & \pc{-1} & \pc{0} & \pc{0} & \pc{0} & \pc{1} & \pc{0} & \pc{1} & \pc{-2} & \pc{-2} & \pc{-1} & \pc{-1} & \pc{-1} & \pc{0} \\
\pc{0} & \pc{0} & \pc{0} & \pc{0} & \pc{0} & \pc{0} & \pc{0} & \pc{0} & \pc{0} & \pc{-1} & \pc{-1} & \pc{-1} & \pc{-1} & \pc{-1} & \pc{-1} \\
\pc{0} & \pc{0} & \pc{-1} & \pc{0} & \pc{0} & \pc{0} & \pc{1} & \pc{-1} & \pc{0} & \pc{-1} & \pc{-1} & \pc{0} & \pc{-1} & \pc{-1} & \pc{0} \\
\end{array}
\right) \\
\outcell{8} & \left(
\begin{array}{ccccccccccccccc}
\pc{0} & \pc{-1} & \pc{-1} & \pc{0} & \pc{-1} & \pc{0} & \pc{0} & \pc{0} & \pc{0} & \pc{-2} & \pc{-2} & \pc{-1} & \pc{-1} & \pc{-1} & \pc{-1} \\
\pc{0} & \pc{0} & \pc{-1} & \pc{-1} & \pc{-1} & \pc{0} & \pc{0} & \pc{-1} & \pc{0} & \pc{-1} & \pc{-2} & \pc{-1} & \pc{-1} & \pc{-2} & \pc{-1} \\
\pc{0} & \pc{0} & \pc{-1} & \pc{0} & \pc{0} & \pc{0} & \pc{1} & \pc{0} & \pc{1} & \pc{-2} & \pc{-2} & \pc{-1} & \pc{-1} & \pc{-1} & \pc{0} \\
\pc{0} & \pc{0} & \pc{0} & \pc{0} & \pc{0} & \pc{0} & \pc{0} & \pc{0} & \pc{0} & \pc{-1} & \pc{-1} & \pc{-1} & \pc{-1} & \pc{-1} & \pc{-1} \\
\pc{0} & \pc{0} & \pc{-1} & \pc{0} & \pc{0} & \pc{0} & \pc{1} & \pc{-1} & \pc{0} & \pc{-1} & \pc{-1} & \pc{0} & \pc{-1} & \pc{-1} & \pc{0} \\
\end{array}
\right) \\
\outcell{9} & \pc{\{}\frac{\pc{X[2,1]X[4,1]X[6,1]}}{\pc{X[1,3]X[1,5]X[1,6]}},\frac{\pc{X[1,3]X[2,3]}}{\pc{X[3,2]X[3,6]}},\frac{\pc{X[1,5]}}{\pc{X[5,4]X[5,6]}},\frac{\pc{X[3,2]X[6,2]}}{\pc{X[2,1]X[2,3]}}, \\
& \; \; \frac{\pc{X[1,6]X[3,6]X[5,6]}}{\pc{X[6,1]X[6,2]X[6,4]}},\frac{\pc{X[1,6]X[2,3]X[5,4]}}{\pc{X[4,1]X[6,2]}}\pc{\}}
\end{flalign*}
There are several features worth highlighting. First, we can easily see that the first loop variable in the basis is the loop around face 1; indeed, it is precisely the first row which is gauged when going from the master space to the moduli space. Secondly, we see that the first perfect matching, i.e.\ \pc{X[1,3]X[1,5]X[1,6]X[2,3]} found using \pc{perfectMatchings}, is the reference matching, by noting that the first column in \pc{master} only contains zeros. For example, it is easy to verify that the third perfect matching divided by \pc{X[1,3]X[1,5]X[1,6]X[2,3]} equals \pc{basis[[2]]}$^{\pc{-1}}$\pc{basis[[3]]}$^{\pc{-1}}$\pc{basis[[4]]}$^{\pc{-1}}$ \pc{basis[[6]]}$^{\pc{-1}}$.

The final thing to note is that this example has two independent closed loops in the variable basis. Hence, evaluating \pc{moduliLoopVariablesBFT} using gauging 2 will result in a moduli space whose coordinates are fewer than those of the master space by two rather than one.
}%
\end{itemize}
As can also be seen by the example above, evaluating \pc{moduliLoopVariablesBFT} using gauging 1 will by default use the variables from \pc{loopVariablesBasis} with the input \var{standardfacevariables} set to \pc{True}, whereas using gauging 2 will use \pc{loopVariablesBasis} with the input \var{standardfacevariables} set to \pc{False}. In either case it will gauge precisely those loop-variables belonging to the \textit{first} element of the list returned by \pc{loopVariablesBasis}.

\pointbft \func{higgsEdgesBFT}{\kast,\varbar{edgelisttohiggs}}: Higgses away edges in \var{edgelisttohiggs} from the Kasteleyn in a consistent way.\footnote{By ``consistent'' we mean according to the output of \pc{consistentEdgeRemoval} in \sref{sec:generalbipartitafunctions}.} The removal of edges will merge faces; this function will automatically relabel the faces to reflect this, and rename certain edges in order to avoid duplicate edge labels. The function \pc{higgsEdgesBFT} will then return a list containing the four components of the Kasteleyn, in the usual order of \pc{\{topleft,topright,bottomleft,bottomright\}}. We note that the renaming of edges works by tagging additional symbols \pc{X} to the edge names; for example, if after the merging of faces there are three edges called \pc{Z[1,2]}, it will rename one of them \pc{XZ[1,2]} and another one \pc{XXZ[1,2]}.\footnote{A word of caution: there is a possibility that the automatically generated edge names conflict with variables the user has previously defined. In such cases it is advisable to use a completely different naming system for the Kasteleyn given to \pc{higgsEdgesBFT} compared to previously defined variable names, e.g.\ by naming edges \pc{X[}$i,j$\pc{]} in one context and \pc{edge[}$i,j$\pc{]} in another.}
\begin{itemize}
\item[] 
{\small \textit{Example.} Using the example given previously to illustrate \pc{moduliLoopVariablesBFT} (where we assume \pc{topleft}, \pc{topright}, \pc{bottomleft} and \pc{bottomright} to be defined in the same way), we illustrate how higgsing edges is easily done with \pc{higgsEdgesBFT}.
\begin{flalign*}
\incell{6} & \pc{higgsEdgesBFT[topleft,topright,bottomleft,bottomright,} \\
& \; \; \pc{\{X[6,2],X[2,1]\}]} \\
\outcell{6} & \pc{\{\{\{0,0,X[1,1],0\},\{XX[3,1],X[1,3],0,XX[1,1]\},\{0,0,0,X[1,5]\}\},} \\
& \; \; \pc{\{\{0\},\{0\},\{X[5,1]\}\},\{\{XX[1,3],0,0,0\},\{0,X[3,1],0,0\},\{0,0,0,0\},} \\
& \; \; \pc{\{0,0,0,XX[5,1]\}\},\{\{0\},\{0\},\{0\},\{0\}\}\}} 
\end{flalign*}
}%
\end{itemize}

\pointbft \func{zigZags}{\kast,\varbardef{invertedrule}{False}}: Returns a list of all zig-zag paths, where each path is expressed as a product of edges. If the edge appears in the numerator it is oriented from white node to black node and if it appears in the denominator it is oppositely oriented. A zig-zag path turns maximally left when arriving at a white node and turns maximally right when arriving at a black node; it is possible to apply the opposite rule by setting the optional input \var{invertedrule} to \pc{True}.

\pointbft \func{zigZagNumeratorsDenominators}{\kast}: Returns a list of all zig-zag paths, which turn left at white nodes and right at black nodes as for the function \pc{zigZags}, albeit expressed differently to \pc{zigZags}: each zig-zag path is now expressed as a sequence of edges it traverses, in chronological order. This is presented as a list of two elements: the first element is a list containing the edges from white nodes to black nodes, in chronological order, and the second  is a list containing the edges from black node to white node, also in chronological order. If a zig-zag path traverses an edge in both directions it will appear in both the first and the second list, contrary to the expression given by the \pc{zigZags} function, where this edge will not appear anywhere.
\begin{itemize}
\item[] 
{\small \textit{Example.} For illustrative purposes it can be useful to compare the expressions given by \pc{zigZags} and \pc{zigZagNumeratorsDenominators} for a simple example embedded on genus zero with two boundaries.
\begin{flalign*}
\incell{2} & \pc{zigZags[\{\{X[2,1],X[1,2]\},\{X[4,2],X[2,3]\},\{X[1,4],X[5,1]\}\},} \\
& \; \; \pc{\{\{X[2,2],0,0\},\{0,X[3,4],0\},\{0,0,X[4,5]\}\},\{\{0,X[3,5]\}\},\{\{0,0,0\}\}]} \\
& \pc{zigZagNumeratorsDenominators[\{\{X[2,1],X[1,2]\},\{X[4,2],X[2,3]\},} \\
& \; \; \pc{\{X[1,4],X[5,1]\}\},\{\{X[2,2],0,0\},\{0,X[3,4],0\},\{0,0,X[4,5]\}\},} \\
& \; \; \pc{\{\{0,X[3,5]\}\},\{\{0,0,0\}\}]} \\
\outcell{2} & \left\{\frac{\pc{X[1,4]X[2,3]}}{\pc{X[4,2]X[5,1]}},\frac{\pc{X[2,1]X[4,5]}}{\pc{X[1,4]X[2,2]}},\frac{\pc{X[1,2]X[4,2]}}{\pc{X[2,1]X[2,3]}},\frac{\pc{X[2,2]X[5,1]}}{\pc{X[1,2]X[4,5]}}\right\} \\
\outcell{3} & \pc{\{\{\{X[3,5],X[1,4],X[2,3]\},\{X[5,1],X[4,2],X[3,5]\}\},} \\
& \; \; \pc{\{\{X[2,1],X[4,5]\},\{X[2,2],X[1,4]\}\},\{\{X[4,2],X[1,2],X[3,4]\},} \\
& \; \; \pc{\{X[3,4],X[2,1],X[2,3]\}\},\{\{X[5,1],X[2,2]\},\{X[4,5],X[1,2]\}\}\}}
\end{flalign*}
As can already be seen from the first zig-zag path, there should be three edges in the numerator and three edges in the denominator, but since \pc{X[3,5]} appears in both lists in the first zig-zag returned by \pc{zigZagNumeratorsDenominators}, the edges cancel in the final expression for the zig-zag path. Moreover, from \pc{zigZagNumeratorsDenominators[[1]]} we see that the order in which the edges are traversed in this zig-zag are \pc{X[3,5]}, \pc{X[5,1]}, \pc{X[1,4]}, \pc{X[4,2]}, \pc{X[2,3]}, \pc{X[3,5]}.
}%
\end{itemize}

\pointbft \func{selfIntersectingZigZagsQ}{\kast}: Self intersecting zig-zag paths are defined as those which cross themselves, which happens when such a path traverses an edge in both directions along its course. The function \newline \pc{selfIntersectingZigZagsQ} returns \pc{True} if the BFT has self-intersecting zig-zag paths and \pc{False} otherwise. 

\pointbft \func{badDoubleCrossingZigZagQ}{\kast}: Two different zig-zag paths may cross each other multiple times; after a pair of zig-zag paths cross, a ``bad double crossing'' is declared to have occurred if the same pair of zig-zag paths cross again in the ``future'' of both zig-zag paths.\footnote{For a more precise definition of a bad double crossing the reader is referred to \cite{ArkaniHamed:2012nw} and \cite{2006math09764P}.} The function \newline \pc{badDoubleCrossingZigZagQ} returns \pc{True} if a bad double crossing occurs for some pair of zig-zag paths and returns \pc{False} otherwise.

\pointbft \func{badDoubleCrossingZigZagPairs}{\kast}: Returns a list where each element contains a pair of zig-zag paths which have a ``bad double crossing''. The zig-zag paths are each expressed as a single list of edges, in chronological order, traversed by the zig-zag path; this view makes the bad double crossing more transparent, since a bad double crossing happens when two edges $X[i]$ and $Y[j]$ appear in both zig-zag paths, and in both cases $X[i]$ appears before $Y[j]$.
\begin{itemize}
\item[] 
{\small \textit{Example.} The example given to illustrate \pc{zigZagNumeratorsDenominators} has a bad double crossing.
\begin{flalign*}
\incell{2} & \pc{badDoubleCrossingZigZagPairs[\{\{X[2,1],X[1,2]\},\{X[4,2],X[2,3]\},} \\
& \; \; \pc{\{X[1,4],X[5,1]\}\},\{\{X[2,2],0,0\},\{0,X[3,4],0\},\{0,0,X[4,5]\}\},} \\
& \; \; \pc{\{\{0,X[3,5]\}\},\{\{0,0,0\}\}]} \\
\outcell{2} & \pc{\{\{\{X[3,5],X[5,1],X[1,4],X[4,2],X[2,3],X[3,5]\},} \\
& \; \; \; \; \pc{\{X[3,4],X[4,2],X[2,1],X[1,2],X[2,3],X[3,4]\}\}\}}
\end{flalign*}
From which we can read off that there is a single bad double crossing, since the returned list has only one element, and the double crossing occurs between the zig-zag paths $\frac{\pc{X[1,4]X[2,3]}}{\pc{X[4,2]X[5,1]}}$ and $\frac{\pc{X[1,2]X[4,2]}}{\pc{X[2,1]X[2,3]}}$. From the  output we can see that \pc{X[4,2]} appears before \pc{X[2,3]} in both zig-zag paths.
}%
\end{itemize}

\pointbft \func{zigZagsAsPerfectMatchings}{\kast,\newline \phantom{444444}\varbardef{invertedrule}{False}}: For a subclass of BFT$_1$s known as dimer models, all zig-zag paths can be expressed as a ratio of perfect matchings.\footnote{This is useful for example when studying the D-brane construction under partial resolution \cite{GarciaEtxebarria:2006aq}, or for constructing larger cluster integrable systems from simpler ones \cite{Franco:2012hv}.} This will not hold for general BFTs, although it is still often possible to express certain zig-zag paths as ratios of perfect matchings. The function \pc{zigZagsAsPerfectMatchings} goes through each of the zig-zag paths given by \pc{zigZags} and gives all pairs of perfect matchings whose ratio equals the expression for the zig-zag path. The function returns a list where each element corresponds to a zig-zag path, expressed as a list of pairs of perfect matchings. It is possible to define zig-zag paths to turn right at white nodes and left at black nodes, instead of the opposite, default choice; changing the direction of the zig-zags in this way is done by setting \var{invertedrule} to \pc{True}.

\pointbft \func{reducibilityBFTQ}{\kast,\varbardef{gauging}{2}}: Returns \pc{True} if it possible to remove some edge without modifying the moduli space, and returns \pc{False} if the removal of any edge will remove points in the toric diagram of the moduli space. The moduli space, and hence the output of \pc{reducibilityBFTQ}, depends on the chosen gauging. The default option for the gauging is to use gauging 2; to check reducibility under gauging 1 it is necessary to set \var{gauging} to \pc{1}.

\pointbft \func{reducibilityBFTedges}{\kast,\varbardef{gauging}{2}}: Returns a list of edges which, if removed, do not affect the moduli space. If no edges may be removed it returns an empty list. As for \pc{reducibilityEdges} in \sref{sec:scatteringinfofunctions}, it is generally not possible to \textit{simultaneously} remove all edges given by \pc{reducibilityBFTedges}. The default gauging used to compute the moduli is gauging 2; to evaluate the moduli space using gauging 1 it is necessary to set the parameter \var{gauging} to \pc{1}.

\pointbft \func{reductionGraphBFT}{\kast,\varbar{gauging}}: Returns a list where each element is a list of edges which may be simultaneously removed without altering the moduli space. Removing all edges in any element of the list returned by \pc{reductionGraphBFT} constitutes a reduction; however, not all reductions given by \pc{reductionGraphBFT} are \textit{maximal} reductions: the function begins with the smallest possible reductions and ends with all maximal reductions. All reductions returned by \pc{reductionGraphBFT} are consistent edge removals, in the sense explained in \pc{consistentEdgeRemoval}.

\section{Advice and Final Words} 
\label{sec:finalwords}

The package provided with this documentation has been designed for ease of use and computational speed. It has been thoroughly tested on millions of examples, including highly idiosyncratic ones. To further ensure a smooth user experience of the package, we here supply a few points of advice.
\begin{itemize}
\item Many functions in the package are requested by other functions, not to mention the user, a large number of times. Some of these functions are also computationally intensive and take a longer time to compute. Therefore, they have been endowed with a ``memory'', namely that they remember the input and the corresponding output, such that if they are called again they avoid recomputing the result and immediately return the remembered output. While this offers a large speed increase, each time the function is called with a different input it will store input and output in the computer's memory.\footnote{If implemented inattentively, the memory property will cause functions to recall values for optional parameters which have changed, thus leading to seemingly peculiar behavior. This is not an issue in this package; it has been avoided by defining optional parameters using a two-stage process.} Fortunately, trading memory for speed is rarely an issue in the package; the trade-off is likeliest to be noticeable when performing a large scan on huge numbers of different models, e.g.\ when scanning over thousands of planar and non-planar on-shell diagrams. The functions in the package can be made memoryless by simply controlling a package-defined global parameter:\newline
$\incell{2} \pc{functionMemory}\;\pc{=}\;\pc{False;}$\newline
which will ensure that the functions stop storing input and output information.\footnote{Conversely, setting \pc{functionMemory} to \pc{True} will again begin storing input and output data.} Because of how the \textit{stratification} functions in \sref{sec:stratificationfunctions} work, this will not cause a noticeable difference in their computational speed.

If functions have already stored a large amount of input and output information, to clear these results from the memory it is possible to use the \pc{CleanSlate} package, or more bluntly by restarting the kernel.
\item The interactive function \pc{drawGraph} updates dynamically, which in Mathematica happens in the Front End, the part responsible for dealing with the notebook, accepting inputs and delivering outputs of computations. The Front End is independent of the kernel---this ensures that it is possible, for example, to scroll in a notebook while a computation is running. It also means that it is possible to use \pc{drawGraph} while other computations are running. Because the Front End is responsible for a variety of tasks, it is important that it does not become entirely absorbed in a complicated calculation in \pc{drawGraph}. Therefore, by default, Mathematica has a time limit on calculations in \pc{drawGraph}: if the evaluation takes longer than 5 seconds, it will  abort the evaluation, which will also cause the drawn graph to disappear. 

The function \pc{drawGraph} has been designed to perform time-consuming tasks in the kernel, and quick operations such as drawing in the Front End. It is theoretically possible, however, to draw graphs of such size and complexity that the Front End cannot cope within the 5-second limit, causing the user to lose the drawn graph. It is good practice, therefore, to use the ``Print Kasteleyn components'' button as a way to save progress on large graphs. Alternatively, the ``Print graph'' button can be used to a similar end.
\item Some functions, e.g.\ \pc{squareMove}, create new edges in the graph. There is a very small chance that the names given to the new edges have already been defined and used elsewhere in the user's code. Even though the edges supplied from \pc{bipartiteSUSY} should belong to a different ``Mathematica context'', and \pc{HoldForm} has been used whenever deemed appropriate, it is still good practice to use a different naming system for user-defined quantities and edges in the diagram, e.g.\ \pc{edge[}$i$\pc{,}$j$\pc{]} in the Kasteleyn and \pc{e[}$i$\pc{,}$j$\pc{]} elsewhere.
\item In rare cases, when working with BFT$_1$ models, the function \pc{squareMove} can struggle to identify the face in which to perform the square move; this has to do with the importance of the embedding surface of  BFT$_1$s,\footnote{The problem only occurs when the square face only involves two rather than four distinct nodes, by looping around the higher-genus surface in particular ways.} and does not occur when dealing with scattering amplitudes. If \pc{bipartiteSUSY} is unsure how to proceed, it will print out a warning. These scenarios are easy to fix: introducing bivalent nodes in the graph such that the square face actually utilizes four distinct nodes will resolves the ambiguity encountered by the program.
\item If the package is loaded after defining symbols which also appear in the package, Mathematica may issue warnings when the package is loaded. To avoid this, it is recommended to always load the package first, before introducing new variables and functions in a Mathematica session.
\item Because of how pattern recognition is used in the package, for BFT$_1$s it is recommended to always label faces, and hence edges, with positive non-zero integers.
\end{itemize}
Finally, we highlight that the code for the package has been designed to be \textit{clean and transparent} in order to facilitate future edits and improvements. To this end, along with the source files on the arXiv containing the package, a notebook version of the package is included, complete with user-friendly indentation and comments in the code. We point out that functions defined in the package have, in reality, additional optional input-variables to those shown in the documentation in \sref{sec:functionlist}. The additional variables can be safely ignored; they are present to make the package more flexible to future enhancements.

The package undoubtedly has scope for further refinement and augmentation; such changes are very welcome and very actively encouraged, provided this package is quoted as the originating platform on which the improvements were made.

The primary goal in the creation of this package was to make intuitive and accessible a wide range of technical knowledge acquired over recent years, and for it to be a stepping stone in the development of yet more advanced methods.

\section*{Acknowledgements}

The creation of this package was prompted by a collection of new results and techniques developed over several years preceding this work. To this end, I would like to thank in particular my collaborators Jacob Bourjaily, Sebasti\'{a}n Franco, Alberto Mariotti, Brenda Penante, Jaroslav Trnka and Congkao Wen for useful discussions along this process. I would also like to thank the Simons Summer Workshop at the Simons Center for Geometry and Physics, where part of the work for this project was performed.


\bibliographystyle{JHEP}
\bibliography{bipartiteSUSY}
\end{document}